# Band-type resonance: non-discrete energetically-optimal resonant states

Arion Pons[1,2,*] and Tsevi Beatus[1,2,†]

[1]*The Benin School of Computer Science and Engineering; The Hebrew University of Jerusalem, Giv'at Ram, Jerusalem, Israel.*

[2]*The Silberman Institute of Life Sciences, The Hebrew University of Jerusalem, Giv'at Ram, Jerusalem, Israel.*

**Abstract:** Structural resonance involves the absorption of inertial loads by a tuned structural elasticity: a process playing a key role in a wide range of biological and technological systems, including many biological and bio-inspired locomotion systems. Conventional linear and nonlinear resonant states typically exist at specific discrete frequencies, and specific symmetric waveforms. This discreteness can be an obstacle to resonant control modulation: deviating from these states, by breaking waveform symmetry or modulating drive frequency, generally leads to losses in system efficiency. Here, we demonstrate a new strategy for achieving these modulations at no loss of energetic efficiency. Leveraging fundamental advances in nonlinear dynamics, we characterise a new form of structural resonance: band-type resonance, describing a continuous band of energetically-optimal resonant states existing around conventional discrete resonant states. These states are a counterexample to the common supposition that deviation from a linear (or nonlinear) resonant frequency necessarily involves a loss of efficiency. We demonstrate how band-type resonant states can be generated via a spectral shaping approach: with small modifications to the system kinematic and load waveforms, we construct sets of frequency-modulated and symmetry-broken resonant states that show equal energetic optimality to their conventional discrete analogues. The existence of these non-discrete resonant states in a huge range of oscillators – linear and nonlinear, in many different physical contexts – is a new dynamical-systems phenomenon. It has implications not only for biological and bio-inspired locomotion systems but for a constellation of forced oscillator systems across physics, engineering, and biology.

## 1. Introduction

Resonance is a phenomenon which is both complex and ubiquitous. In a classical mechanics context, structural resonance represents both a harmonic coupling effect between structural inertia and structural elasticity; and a fundamental characteristic of linear oscillators and linear operators [1–3]. This same characteristic generalises to many forms of nonlinear oscillator, and extends to a wide range of other physical contexts: fluid mechanics [4–6], optomechanics [7–10]; electrical circuits [11, 12], plasma dynamics [13], and further [14–17]. Nonlinear resonant phenomena come in many forms, and typically generalise certain optimality properties of linear resonance: optimality in amplitude response [14, 15, 18–20], optimality in energy transfer [20–

---

\* arion.pons@mail.huji.ac.il

† tsevi.beatus@mail.huji.ac.il

25], *etc.* Even focusing on a structural or mechanical context, linear and nonlinear resonance are of core importance to many technological and biological systems: microelectromechanical (MEMS) oscillators [26–28]; atomic force microscopes [29, 30] elastic metamaterials of various forms [31, 32]; the locomotive structures of a huge range of airborne [33–39], aquatic [40–45], and legged organisms [46]; and a correspondingly wide range of bio-inspired locomotive technologies [47–55].

Resonant states are typically discrete, in some sense: linear second-order oscillators have certain discrete eigenstates, and discrete transfer function peaks [1–3]. These linear resonant states exist at certain discrete frequencies, and harmonic input/output waveforms. Nonlinear resonant states – of which there are several different forms – can show more complex behaviour: for instance, a dependency of frequency on input/output amplitude, as in the Duffing oscillator [15, 18, 56]. Nevertheless, these nonlinear resonant states are still discrete in the sense that, at a given amplitude, certain discrete resonant frequencies exist. The discreteness of linear and nonlinear resonant states can be an obstacle to system control. Consider, for instance, classical-mechanical resonant behaviour within biological locomotive systems. If the nominal state of locomotion is resonant, then it should be impossible to modulate the frequency of the locomotive output response, or break the symmetry of the output waveform, without the penalty of leaving the state of resonance – namely, without compromising on efficiency and perfect inertial load absorption [36, 37].

The apparent restrictiveness of biological locomotive resonance is all the more interesting because it is defied in practice. Several species of insect which are thought to utilise resonance to generate a propulsive wingbeat response do engage in significant wingbeat frequency modulation – in both a controlled [36, 57–60], and seemingly stochastic [61] manner. Symmetry-breaking modulation is also observed: in these species, orientation control is regularly achieved via symmetry-breaking modulations in the wingbeat response [62–64]. Both frequency-modulation and symmetry-breaking control appear inconsistent with maintaining a resonant state in a linear oscillator with fixed structural properties: they represent changes in the equilibrium point and natural frequency of the oscillator, respectively. In insects, the operation of sets of auxiliary flight muscles have been proposed as a mechanism for real-time control of thoracic structural resonant properties [36, 38, 65–68] – potentially allowing wingbeat frequency and symmetry-breaking modulation while maintaining a resonant state. Alternatively, this modulation has been explained in terms of the trade-off between deviating from resonance, and achieving other goals, such as performing a harsh manoeuvre [36], or engaging in courtship [69]. This trade-off could involve switching between different system resonant frequencies (displacement, velocity resonance, *etc.*); as well as the particular behaviour of energy resonance, *i.e.*, global resonance [20–25], which shows a particularly low cost associated with frequency deviation [20]. In addition, nonlinear (*e.g.*, Duffing-type) effects could lead to certain forms of amplitude-coupled resonant frequency modulation [20, 34]. Aspects of these proposed explanations are still unclear – for instance, given that wingbeat amplitude modulation (which maintains linear resonance) is available for apparently the same purpose as frequency modulation (which disrupts linear resonance), it is not clear why the latter should ever have evolved [36, 38, 57].

In this work we offer more clarity on this practical topic, by deriving and demonstrating a new dynamical-systems phenomenon. We show how resonant states can be non-discrete: in both linear and nonlinear systems, an energy resonant state [20–25] available not only at discrete eigenstates, but within a band of frequencies and symmetry-broken waveforms around these states. These non-discrete states – the *band-type resonant states* – generalise the energetic properties of a linear eigenstate to non-harmonic input/output waveforms: in practical, classical-mechanical terms, they ensure that the system's inertial power requirements are completely absorbed by elastic effects. We demonstrate how these band-type resonant states can be generated via spectral shaping processes: small modifications to the system drive waveforms can be utilised to generate targeted resonant modulations in frequency or output symmetry. We demonstrate both analytical and numerical methods for carrying out this spectral shaping process – methods which lead to a suite of new control principles for linear and nonlinear systems.

This principle of band-type resonance applies to a wide range of linear and nonlinear oscillators, in many different physical contexts. It represents a fundamental advance in linear and nonlinear dynamics: in our understanding of the relationship between energy transfer and mechanical resonance; and in our ability to control resonant systems. In the context of insect flight motor resonance, band-type resonance quantitatively explains resonant frequency variation effects observed in several insect species. In the context of bio-inspired locomotion systems, it provides new control principles for energetically optimal modulation of locomotive output frequency and symmetry. And in a wider context, it establishes a new subfield of linear and nonlinear dynamics, dealing with energetically-optimal resonant state control in a wide range cutting-edge technologies based on forced oscillation.

## 2. Dynamics of PEA and SEA systems

### 2.1. System configurations

Consider a general single-degree-of-freedom (1DOF) system, with general nonlinear time-invariant dynamics. The system is driven by an actuator, which provides some force, $F(t)$, and thereby generates a system displacement, $x(t)$. In general, we are interested in cases in which $F(t)$ and $x(t)$ are oscillatory. The equation of motion of this system is given by:

$$D(x, \dot{x}, \ddot{x}, \dots) = F(t), \tag{1}$$

where $D(\cdot)$ expresses the system's time-invariant dynamics. Eq. 1 can be used to model a huge range of different physical processes: not only mechanical systems, such as insect flight motors [20] and MEMS devices [28]; but also, electrical [11, 12], and other [15, 17] processes.

Consider then introducing a conservative force into Eq. 1: a force, $F_s(x)$, dependent solely on displacement, or, in structural terms, an elastic element. Introducing this elastic element can lead to two distinct system configurations: parallel-elastic actuation (PEA), in which the actuator and elastic element are connected in parallel; and series-elastic actuation (SEA), in which the actuator and elastic element are connected in series (Fig. 1A,B). The equations of motion of these two configurations may be expressed:

$$\begin{aligned} \text{PEA:} \quad & D(x, \dot{x}, \ddot{x}, \dots) + F_s(x) = F(t), \\ \text{SEA:} \quad & D(x, \dot{x}, \ddot{x}, \dots) = F_s(u - x) = F(t), \end{aligned} \tag{2}$$

where $F_s(x)$ is the conservative force, which may show a nonlinear dependence on $x$; and $u(t)$ is the displacement of the system actuator, or actuation point, which in the SEA configuration may be distinct from $x(t)$. We note that the system dynamics, $D(\cdot)$, may already express some conservative force, which in the PEA case may function identically to $F_s(x)$. The choice between PEA and SEA configurations is a design question found frequently in robotics [70, 71], and a distinguishing factor between many forms of physical system [20, 28].

To analyse the oscillatory behaviour of these general nonlinear systems, we use work-loop analysis techniques [25]. First, consider that $x(t)$, the system motion, is given (prescribed, or desired) as some periodic (*i.e.*, steady-state) wave. Then, in the PEA system, it is possible to compute $F(t)$, which is also periodic, and visualise it against $x(t)$: a visualisation as a steady-state work loop, in the plane $F$-$x$. We require, as conditions defining the limits of our analysis, that this steady-state work loop: (**i**) is a closed simple curve (*i.e.*, no self-intersection); (**ii**) is no more than bivalued at any $x$; and (**iii**) represents net power dissipation (*i.e.*, the progression of time must represent clockwise travel around the closed loop). A corollary of condition (**ii**) is the simpler condition that the given periodic $x(t)$ must be composed of two half-cycles that are each monotonic. Under these conditions, the steady-state work loop ($F$ vs. $x$) permits a representation as two single-valued curves: $F^+(x)$ and $F^-(x)$ (Fig. 1A, with examples in appendix A.1). Correspondingly, we could do the same for the system the inelastic load, $G(t) = D(x, \dot{x}, \ddot{x}, \dots)$, yielding $G^+(x)$ and $G^-(x)$, and the work-loop form of the PEA system:

$$\text{PEA:} \quad F^\pm(x) = G^\pm(x) + F_s(x). \tag{3}$$

The work-loop formulation of an SEA system is analogous. Again, consider that $x(t)$ is given as some periodic wave. Then, we may directly compute $F(t)$, which is also periodic, and visualise the system work-loop in the (rotated) $x$-$F$ plane. We again require, as analysis conditions, that this steady-state work loop: (**i**) is a closed simple curve; (**ii**) is no more than bivalued at any $F$; and (**iii**) represents net power dissipation (*i.e.*, the progression of time must represent counterclockwise travel around the closed loop). Again, a corollary of condition (**ii**) is the simpler condition that the computed periodic $F(t)$ is composed of two half-cycles that are each monotonic (*NB*: a condition on $F(t)$ which can be satisfied even when $x(t)$ does not satisfy this same condition). The steady work loop ($x$ vs. $F$) then permits a representation as two single-valued curves: $X^+(F)$ and $X^-(F)$ (Fig. 1B, and appendix A.1). We can do the same to the curves of $u$ vs. $F$, and represent them as $U^+(F)$ and $U^-(F)$. This would yield the work-loop form of the SEA system:

$$\text{SEA:} \quad U^\pm(F) = X^\pm(F) + F_s^{-1}(F), \tag{4}$$

where, $F_s^{-1}(\cdot)$ is the inverse of $F_s(\cdot)$ . In the SEA case, we assumed that the elastic profile, $F_s(\cdot)$, is invertible – precluding, for instance, bistable elasticities. The analogy between Eqs. 3-4 may be noted: both represent the sum of an inelastic work loop ($G^\pm(x)$, $X^\pm(F)$) and an elastic profile ($F_s(x)$, $F_s^{-1}(F)$), yielding an elastic work loop ($F^\pm(x)$, $U^\pm(F)$). This form is of key importance to our analysis.

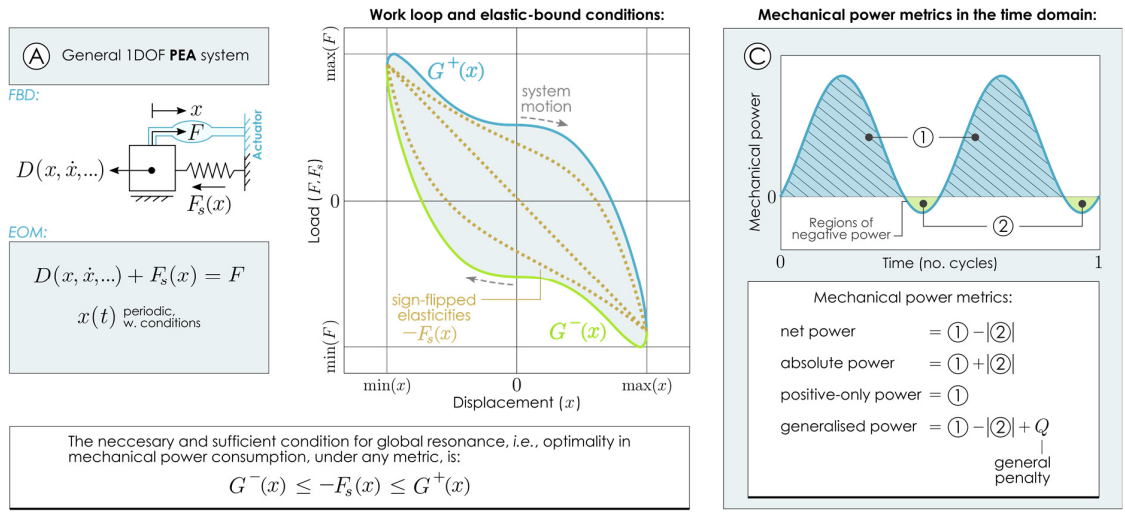

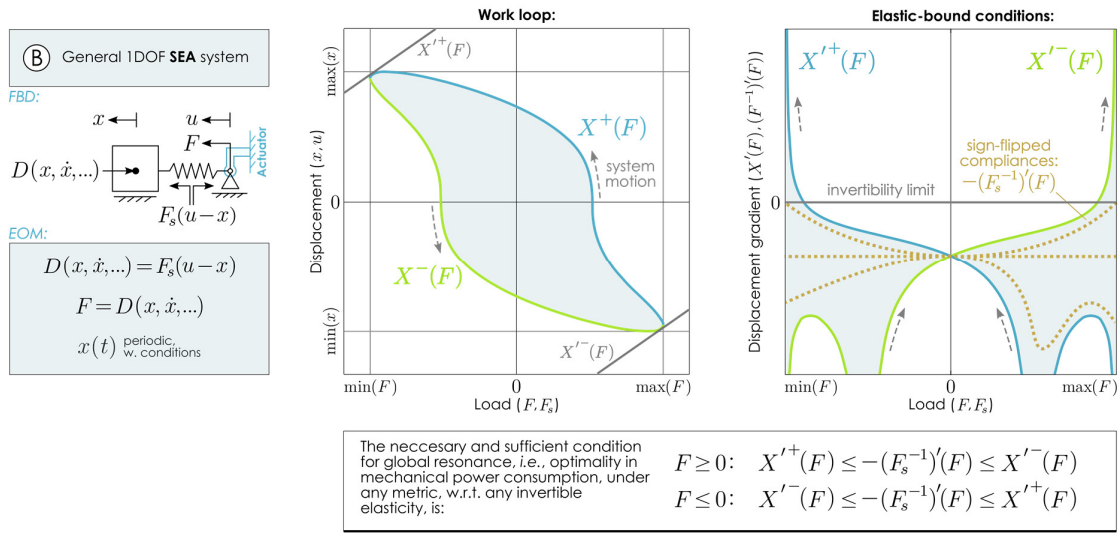

**Fig 1: Schematic of general PEA and SEA systems, and their respective elastic-bound conditions**. (**A**) A general nonlinear 1DOF PEA system, with the elastic-bound conditions for energy resonance illustrated on the system work-loop (Eq. 3). (**B**) A general nonlinear 1DOF SEA system, with the system work-loop illustrated in the rotated work-loop space (Eq. 4), and the elastic-bound conditions for energy resonance illustrated in the relevant gradient space (Eq. 6). (**C**) Explanation of the four metrics of mechanical power consumption – net, absolute, positive-only and generalised, as per Eq. 5 – with a representative time-domain power requirement.

## 2.2. Energy resonance

The work-loop forms of Eq. 3-4 allow a characterisation of the state of energy resonance in these general PEA and SEA systems. Energy resonance, or global resonance, is a form of linear and nonlinear resonance that is closely related to mechanical power consumption [20–25]. In intuitive terms, the state of energy resonance is a state in which power flows only from the actuator to the system, and never in reverse: in a linear PEA 1DOF system, the energy resonant frequency coincides with the system natural frequency, but in other systems, these frequencies may differ [20]. In more formal terms, the time-domain actuator mechanical power consumption, $P(t)$, is given by $P = F\dot{x}$ in the PEA system, and $P = F\dot{u}$ in the SEA system. $P(t)$, may contain regions of both positive ($P > 0$) and negative ($P < 0$) power: positive power represents power flow from the actuator to the system; and negative power, the reverse. Energy resonant states are those in which $P(t) \geq 0$, $\forall t$.

Recent results [25] have shown that energy resonance can be seen not only in terms of power flow, but in terms of power minimisation: energy resonant states form the solution to a particular power minimisation problem. Consider a nonlinear 1DOF PEA or SEA system, undergoing prescribed periodic motion, $x(t)$, of period $T$. If the system elasticity, or conservative force, $F_s(\cdot)$, could be chosen, then what choice(s) of elasticity would minimise the overall mechanical power consumption associated with this prescribed motion? This question is formulated as an inverse problem: finding $F_s(\cdot)$ for a given $x(t)$ that minimises power consumption. The overall power consumption of a system can be expressed as an integral of $P(t)$, but the details of this integral are actuator-specific, depending on how the actuator behaves under negative power. [72–74]. In general, four classes of metrics, $\overline{P}_{(\cdot)}$ can be identified [25]:

$$\text{The net power:} \qquad \overline{P}_{(a)} = \frac{1}{T}\int_0^T P(t)\,dt,$$

$$\text{The absolute power:} \qquad \overline{P}_{(b)} = \frac{1}{T}\int_0^T |P(t)|\,dt,$$

$$\text{The positive-only power:} \qquad \overline{P}_{(c)} = \frac{1}{T}\int_0^T P(t)[P(t) \geq 0]_{\mathbb{I}}\,dt, \qquad (5)$$

$$\text{The generalised power:} \qquad \overline{P}_{(d)} = \frac{1}{T}\int_0^T \left(P(t) + Q(t)|P(t)|[P(t) \leq 0]_{\mathbb{I}}\right)dt,$$

where $[\,\cdot\,]_{\mathbb{I}}$ is the Iverson bracket, with $[\lambda]_{\mathbb{I}} = 1$ for a true statement $\lambda$, and $[\lambda]_{\mathbb{I}} = 0$ for a false statement $\lambda$ [75]; and $Q(t)$ is a general penalty function, $Q(t) > 0$, $\forall t$, penalising negative power. Fig. 1C illustrates these metrics on a representative $P(t)$. For example, if the actuator can absorb all negative power for later use, *i.e.*, the actuator shows perfect energy regeneration [70, 76], then metric $\overline{P}_{(a)}$ is appropriate. If the actuation system shows no regeneration, and must expend energy to account for negative power (such as in a rocket reaction engine), then metric $\overline{P}_{(b)}$ should be used. If the actuation system shows imperfect regeneration, and can dissipate but not store negative power (such as in car brakes), then the relevant metric is $\overline{P}_{(c)}$. Different disciplines of study have conventionally used different $\overline{P}_{(\cdot)}$. The absolute power is used in robotics [77, 78] and bipedal biomechanics [79–82]; the positive-only power, in insect flight energetics [72, 83, 84]; and the generalised power, in models of human metabolic cost [85–87]. The net power $\overline{P}_{(a)}$ is independent of the system elasticity, $F_s(\cdot)$, because elastic forces are conservative. The question of minimising $\overline{P}_{(a)}$ with respect to elasticity is thus irrelevant.

The net power, absolute power, and positive-only power are particular forms of the generalised power, with constant penalty functions $Q = 0$, $Q = 2$, and $Q = 1$, respectively.

Fascinatingly, it is possible to devise a solution for the optimal elasticity, $F_s(\cdot)$, that is simultaneously valid for all three power consumption metrics, $\overline{P}_{(b)-(d)}$. It can be demonstrated that any energy resonant state, *i.e.*, any state in which negative power is absent ($P(t) \geq 0$, $\forall t$) is necessarily a state in which all of these metrics are simultaneously minimised with respect to elasticity. The energy resonance condition then leads to the following explicit solution for the energetically-optimal elasticity, $F_s(\cdot)$,:

> For PEA systems:
> $$G^-(x) \leq -F_s(x) \leq G^+(x), \quad \forall x \in [\min x(t), \max x(t)],$$
>
> For SEA systems:
> $$X'^-(F) \leq -(F_s^{-1})'(F) \leq X'^+(F), \quad \forall F \in [\min F(t), 0],$$
> $$X'^+(F) \leq -(F_s^{-1})'(F) \leq X'^-(F), \quad \forall F \in [0, \max F(t)],$$

(6)

as illustrated in Fig. 1A-B. $(\cdot)'$ denotes $d(\cdot)/dF$. These are the elastic-bound conditions: they are sufficient conditions on the elasticity, $F_s(\cdot)$, to ensure that all three metrics of actuator power consumption are minimised with respect to $F_s(\cdot)$; and are necessary conditions in a wide class of systems [25]. In general, these conditions define not a single optimal elasticity, but a bounded range of optimal elasticities – one of which may be linear, but the rest are nonlinear (Fig. 1A-B). Each optimal elasticity is associated with a particular input load waveform, $F(t)$: under this input loading, the system generates the prescribed output response, $x(t)$, and operates in energy resonance.

## 2.3. Composite metrics and principles of invariance

The elastic-bound conditions, Eq. 6, define energy resonant states, in $F_s(\cdot)$: states which minimise the system mechanical power consumption, $\overline{P}_{(b)-(d)}$, at some given state of periodic kinematic output, $x(t)$. Significantly, however, these energy resonant states do not *only* optimise $\overline{P}_{(b)-(d)}$, but also range of other physically-relevant power consumption and performance metrics. The behaviour of some such metrics can be viewed through the lens of two principles of invariance [25]:

**First principle of invariance.** In any PEA system with periodic kinematic output, $x(t)$, the actuator load profile, $F(t)$ is in general dependent on the system elasticity, whereas the actuator displacement profile, $x(t)$, is independent of this elasticity – it is prescribed. Hence, any metric which is a functional of solely $x(t)$ (*e.g.*, $\max x$) is invariant with respect to elasticity. Turning to any SEA system, we observe the reverse: at some given state of periodic kinematic output, $x(t)$, the actuator displacement profile, $u(t)$ is in general dependent on the system elasticity, whereas the actuator load profile, $F(t)$, is always independent of this elasticity – it is prescribed by $F = D(x, \dot{x}, \ddot{x}, ...)$ (Eq. 2). It follows that in SEA systems, any functional of solely $F(t)$ is invariant with respect to elasticity. If we define a composite cost metric as a sum of a mechanical power consumption metric ($\overline{P}_{(b)-(d)}$) with *any* other functionals of solely $F(t)$, then the only term of this composite cost metric that varies with elasticity, $F_s(\cdot)$, is the mechanical power consumption component. The optimal elasticities for this SEA system, in this composite metric, are thus given again by the elastic bound conditions (Eq. 6).

The key relevance of this first principle of invariance, in SEA systems, is that a wide range of practical cost metrics are indeed functionals of solely $F(t)$ [81]. Eq. 7 defines several: the peak load, $\hat{F}$, which is a metric of actuator capability requirement; the load-squared metric, $\overline{P}_{F^2}$, which is a metric of electrical power consumption [88, 89]; and absolute load metric, $\overline{P}_{|F|}$, which is a metric of muscular power consumption [74, 80–82].

Peak load: $\qquad\qquad\qquad \hat{F} = \max_t F(t).$

Load-squared metric: $\qquad \overline{P}_{F^2} = \frac{1}{T} \int_0^T F(t)^2\, dt,$ $\qquad\qquad$ (7)

Absolute load metric: $\qquad \overline{P}_{|F|} = \frac{1}{T} \int_0^T |F(t)|\, dt,$

The first principle of invariance represents an analogy between PEA and SEA systems, and allows the elastic-bound conditions, originally derived for mechanical-power minimisation, to be extended to a wider class of optimisation problems.

**Second principle of invariance.** As noted in Section 2.3.1, in PEA systems, the actuator load profile, $F(t)$ is dependent on the system elasticity, and thus, so too are functionals of solely $F(t)$, such as the load-squared metric, $\overline{P}_{F^2}$, or the peak load, $\hat{F}$. However, the absolute load metric, $\overline{P}_{|F|}$, is an exception. This metric is invariant with respect to elasticities satisfying the elastic-bound conditions (Eq. 6), provided that the prescribed kinematic waveform, $x(t)$, is composed of two symmetric half-cycles [25]. That is, provided that $x(t) = x(T - t), \forall t$, which is the case for many relevant problems. In general[1], this second principle of invariance means that in PEA systems the elastic-bound conditions, Eq. 6, are also the conditions for an elasticity, $F_s(\cdot)$, to be optimal in any combination of mechanical power $\overline{P}_{(b)-(d)}$ and absolute load ($\overline{P}_{|F|}$). Thus, while in general SEA systems are far superior in optimising useful composite cost metrics, for muscular actuators, PEA systems receive an unexpected advantage, bringing such systems to nearly the same level of optimality as SEA systems. This has significant implications, *e.g.*, for biomimicry: indicating that principles of optimality in muscular systems may not translate to principles of optimality in, *e.g.*, electromechanical systems.

## 3. Band-type resonance: theoretical basis

### 3.1. Definition of band-type resonant states
The formulation of energy resonance outlined in Section 2 represents a synthesis of two existing nonlinear analysis strategies: the global resonance analysis of Ma and Zhang [21–24], and the power optimisation approach of Pons and Beatus [20, 25]. Thus far, however, these energy resonant states have been expressed in terms of a set of optimal elasticities, $F_s(\cdot)$, that ensure energy resonance for a particular output, $x(t)$ – rather than the set of inputs/outputs that are energy resonant in a system with fixed elasticity. The non-uniqueness in the set of optimal elasticities gives us reason to pause for a moment: surely, if these optimal elasticities is non-unique, then the input/output energy resonant state should be non-unique? Consider the general 1DOF PEA and SEA systems described in Section 2.1, now, with some fixed elastic load profile, $F_s(\cdot)$, linear or nonlinear. If, for one such system, some output kinematic waveform, $x(t)$, is prescribed, then we may test whether the relevant elastic-bound conditions (Eq. 6) are satisfied – that is, whether this fixed elasticity, $F_s(\cdot)$, represents one of the optimal elasticities for $x(t)$. If a set of kinematics $\mathcal{X} = \{x_i(t)\}_i$ exists such that a single $F_s(\cdot)$ lies within the common bounded work-loop area of $\mathcal{X}$, then $F_s(\cdot)$ is capable of minimising actuator power consumption for the entire set $\mathcal{X}$. In this scenario, illustrated in Fig. 2, each $x_i(t) \in \mathcal{X}$ is associated with specific actuator waveform(s): for PEA systems, the load waveform $F_i(t)$; and for SEA systems, both the displacement $u_i(t)$ and load $F_i(t)$ waveforms, which are mutually dependent. All of these states are energy resonant: they contain no negative power ($P(t) \geq 0$, $\forall t$), and thus, ensure that power required to generate each particular $x_i(t)$ is the minimum possible power required, given any possible choice of system elasticities. By controlling $u_i(t)$ and/or $F_i(t)$, it is possible to switch between output kinematics, $x_i(t) \in \mathcal{X}$ without leaving an energy resonant state: the system's fixed elasticity is always energetically optimal for each output.

---

[1] Indeed, it appears the case that the value of $\overline{P}_{|F|}$ which is invariant over the elastic-bound region is necessarily the global minimum of $\overline{P}_{|F|}$ over all possible elasticities – though this point deserves further theoretical treatment.

More formally, consider a general output waveform, $x(t)$, parameterised as a functional of period $T$, amplitude $\hat{x}$ and offset, $x_0$, and basis waveform $w(\tau)$ (normalised amplitude, period, and zero mean), in the form $x(t) = x[\hat{x}, T, x_0, w(\tau), t] = \hat{x}w(t/T) + x_0$. Consider also a PEA system (Eq. 3), with dynamics $D(x, \dot{x}, \ddot{x}, \dots)$ and fixed $F_s(x)$. In this system, the waveform $x(t)$ generates an inelastic actuator load requirement $G(t) = D(x, \dot{x}, \ddot{x}, \dots)$, and an elastic (*i.e.*, overall input) load requirement, $F(t)$, via $F(t) = G(t) + F_s(x(t))$. The inelastic load requirement, $G(t)$, generates a work loop $G$-$x$. This work loop may be parameterised $G^{\pm}(x(t))$ if the analysis conditions (Section 2.1) are met. Now, given $\hat{x}$, $T$, $x_0$, and $w(\tau)$, and thus given $x(t)$, we test the following condition:

$$\text{is} \quad G^-\big(x(t)\big) \leq -F_s\big(x(t)\big) \leq G^+\big(x(t)\big), \ \forall t \quad ? \tag{8}$$

If this condition is satisfied for the given $\hat{x}$, $T$, $x_0$, and $w(\tau)$, then the PEA system's fixed elasticity, $F_s(x)$, is energetically optimal for the waveform, $x(t) = x[\hat{x}, T, x_0, w(\tau), t]$. The set of particular waveforms, $\mathcal{X} = \{x_i(t)\}_i$, such that (**i**) the analysis conditions were met, and (**ii**) Eq. 8 is satisfied, we define to be a space of *band-type resonant outputs* for this specific PEA system (with fixed $F_s(x)$, *etc.*). The set of function tuples, $\mathcal{S} = \big\{\{x_i(t), F_i(t)\}\big\}_i$, including both particular output waveforms $x_i(t) \in \mathcal{X}$ and associated input loads $F_i(t)$, we define to be the system's set of *band-type resonant states*, as illustrated in Fig. 2.

Alongside, consider an SEA system (Eq. 4), with dynamics $D(x, \dot{x}, \ddot{x}, \dots)$ and fixed $F_s(\cdot)$. In this system, the general waveform $x(t) = x[\hat{x}, T, x_0, w(\tau), t] = \hat{x}w(t/T) + x_0$ generates a load requirement $F(t) = D(x, \dot{x}, \ddot{x}, \dots)$, and an actuator displacement requirement $u(t) = x(t) + F_s^{-1}\big(F(t)\big)$. The load requirement $F(t)$ generates an inelastic work loop $x$-$F$. This work loop may be parameterised $X^{\pm}\big(F(t)\big)$ if the analysis conditions (Section 2.1) are met. Given $\hat{x}$, $T$, and $x_0$, and thus $x(t)$, we test the following condition:

$$\begin{aligned} \text{is} \ &X'^-\big(F(t)\big) \leq -(F_s^{-1})'\big(F(t)\big) \leq X'^+\big(F_i(t)\big), \ \ \forall F \in \big[\min\big(F(t)\big), 0\big], \text{and} \\ &X'^+\big(F(t)\big) \leq -(F_s^{-1})'\big(F(t)\big) \leq X'^-\big(F_i(t)\big), \ \ \forall F \in \big[0, \max\big(F(t)\big)\big] \ ? \end{aligned} \tag{9}$$

If this condition is satisfied for the given $\hat{x}$, $T$, $x_0$, and $w(\tau)$, then the SEA system's fixed elasticity, $F_s(\cdot)$, is energetically optimal for the waveform, $x(t) = x[\hat{x}, T, x_0, w(\tau), t]$. The set of particular waveforms, $\mathcal{X} = \{x_i(t)\}_i$, such that (**i**) the analysis conditions were met, and (**ii**) Eq. 9 is satisfied, is defined to be the set of *band-type resonant outputs* for this specific SEA system (again, with fixed $F_s(x)$, *etc.*). The set of function tuples, $\mathcal{S} = \{\{x_i(t), u_i(t), F_i(t)\}\}_i$, including particular output waveforms $x_i(t) \in \mathcal{X}$, associated input displacements $u_i(t)$, and associated input loads $F_i(t)$, is this system's set of *band-type resonant states*.

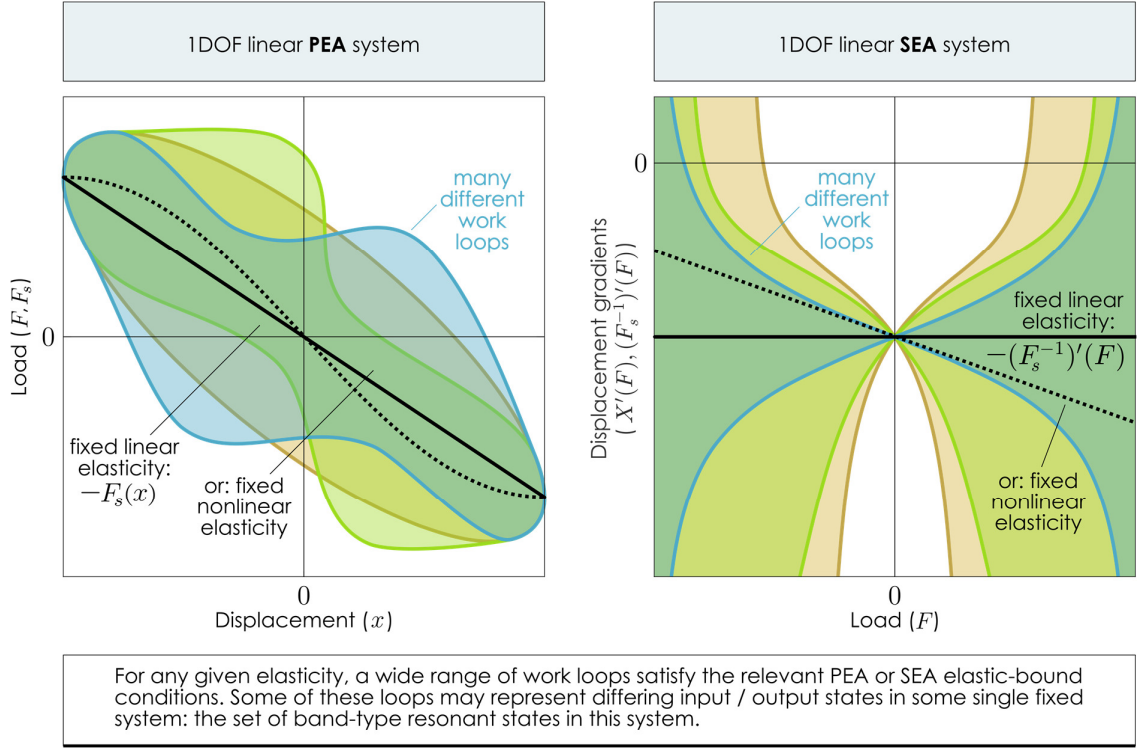

For any given elasticity, a wide range of work loops satisfy the relevant PEA or SEA elastic-bound conditions. Some of these loops may represent differing input / output states in some single fixed system: the set of band-type resonant states in this system.

**Fig 2: Illustration of the principle of band-type resonance.** For any given elasticity, the SEA and PEA elastic-bound conditions for energy resonance are satisfied by a wide range of work loops. If, in a single fixed system, we can find multiple kinematic outputs, $x_i(t)$, and associated inputs, $F_i(t)$, $u_i(t)$, *etc.*, such that the relevant elastic-bound conditions are satisfied, then all of these input/output states are band-type resonant states in this system: at each state, the system's fixed elasticity optimally absorbs inertial power requirements.

### 3.2. Constructing band-type resonant states

We present a few intuitive heuristics to illustrate the properties of these band-type resonant states. For instance: if we start with a particular band-type resonant output for some given system, $x_i(t) \in \mathcal{X}$, then, in *general*, outputs with changes in a single scalar waveform parameter (*i.e.*, a change in only one of the amplitude, period, or offset) will not be band-type resonant outputs. That is to say, in a given system, if $x(t) = \hat{x}w(t/T) + x_0 \in \mathcal{X}$, then, in general, the following $y(t) \notin \mathcal{X}$:

$$
\begin{aligned}
y(t) &= \hat{y}w_i(t/T) + x_0, && \forall \hat{y} \neq \hat{x}, \\
y(t) &= \hat{x}w_i(t/Y) + x_0, && \forall Y \neq T, \\
y(t) &= \hat{x}w_i(t/T) + y_0, && \forall y_0 \neq x_0.
\end{aligned}
\tag{10}
$$

There are exceptions to this heuristic (*e.g.*, amplitude modulation in a linear resonant system). But its validity for most other practical cases can be seen from the following considerations.

The inequality conditions in Eqs. 8-9 include two equality conditions, or, constraints. In the PEA system, this is the condition for load matching at the displacement extrema ($x(t) = \hat{x} \pm x_0$), and in the SEA system, this is the condition for gradient matching at zero load ($F = 0$):

$$\text{PEA:} \quad G^{\pm}(\hat{x} + x_0) = -F_s(\hat{x} + x_0), \qquad G^{\pm}(\hat{x} - x_0) = -F_s(\hat{x} - x_0),$$
$$\text{SEA:} \quad X'^{\pm}(0) = -(F_s^{-1})'(0). \tag{11}$$

These equality conditions must be satisfied in order for $x(t) = \hat{x}w(t/T) + x_0$ to be a band-type resonant output, $x(t) \in \mathcal{X}$. We may observe that $y(t)$ according to Eq. 10 will, in general, not satisfy these equality conditions. For instance, in a linear PEA system, univariate changes to the waveform period, $T$, will break load matching (*e.g.*, because the load at the displacement extremum is the peak inertial load, scaling by $\propto 1/T^2$). Univariate changes to the offset $x_0$ will leave the peak inertial loads invariant, but cause the displacement extrema, $\hat{x} \pm x_0$, to shift, altering the values of the elastic load, $F_s(\hat{x} \pm x_0)$. For example, in any 1DOF system with linear inertia, dissipation, and elasticity, if a particular waveform is resonant (*e.g.*, a simple-harmonic wave) at a particular frequency and zero offset, then this identical wave will not be resonant at different frequencies, or different offsets. This is precisely the mystery regarding insect flight control – *i.e.*, wingbeat frequency [36, 57] and offset [62, 63] modulation – noted in Section 1. The same principle extends to a wide range of nonlinear systems.

Broadly, then: given a single band-type resonant state $x_1(t)$, constructing another different band-type resonant state, $x_2(t)$, requires either varying more than one scalar waveform parameter (amplitude, frequency and offset) and/or varying the basis waveform $w(\tau)$. That is, if we desire a set of band-type resonant states, *e.g.*, over a range of different output waveform periods, $T$, then one option is to find an associated set of different normalised waveforms, $w(\tau)$. This leads to our method for *constructing* band-type resonant states. We first seek a relationship between a scalar waveform parameter (*e.g.*, $T$) and the normalised waveform, $w(\tau)$, such that the system-relevant equality condition (Eq. 11) is satisfied over a range of $T$ (ideally, $\forall T$). This is the process of spectral shaping, and we will describe several analytical and numerical methods for performing it. When we have obtained this relationship, we then assess whether these $\{T, w(\tau)\}$ tuples satisfy the system-relevant inequality condition, Eq. 8 or Eq. 9. We find, by nature of their inequality, that these conditions are satisfied within a certain range (or band) of $T$. Performing this with $T_i$ and $w_i$, we obtain *frequency-band resonance*: energy resonance over a band of frequencies. Performing this with $\hat{x}, x_0$ (in some combination) and $w_i$, we obtain *offset-band resonance*: energy resonance over a band of offset parameters.

Among the huge space of possible band-type resonant effects, frequency-band and offset-band resonance are the two cases that we demonstrate in detail below. These particular cases have significant relevance to many practical systems – including several forms of biomimetic propulsion system. They allow frequency (*i.e.*, thrust magnitude) and offset (*i.e.*, thrust-vectoring) control, while maintaining resonant power optimality. We will demonstrate three different techniques for spectral shaping (obtaining a relationship, *e.g.*, between $T_i$ and $w_i$) that satisfy the system-relevant inequality condition: analytically tuning a generative ordinary-differential equation (ODE); analytically tuning a multi-harmonic waveform; and numerically tuning a multi-harmonic waveform.

## 4. Band-type resonance: applications

### 4.1. Frequency-band resonance in a PEA system, via analytical tuning of a generative ordinary-differential equation (ODE)

As a specific demonstration of frequency-band resonance, consider a 1DOF PEA system (Fig. 1) with fixed linear inertia, linear dissipation, and linear elasticity:

$$
\begin{array}{lll}
\text{Inelastic system dynamics:} & D(\dot{x}, \ddot{x}) = m\ddot{x} + c\dot{x}, & (m > 0, c > 0). \\
\text{Inelastic load requirement:} & G(t) = D(\dot{x}, \ddot{x}) = m\ddot{x} + c\dot{x}. \\
\text{Fixed elasticity:} & F_s(x) = kx, & (k > 0). \\
\text{Elastic load requirement:} & F(t) = G(t) + F_s(x) = m\ddot{x} + c\dot{x} + kx. \\
\text{Natural frequency:} & \omega_0 = \sqrt{k/m}. \\
\text{Damping ratio:} & \zeta = c \,/\, 2\sqrt{km}.
\end{array}
\tag{12}
$$

This linear PEA system can be used as a simple model of insect flight motors [20]; low-Reynolds number [90] and surface-interaction [91] propulsion devices; and MEMS devices [28]. To demonstrate frequency-band resonance in this system, via an analytical spectral shaping process, we use a triangle-like waveform based on that used by Dickinson *et al.* [92] and experimentally verified by others [93] as a model of the wing stroke-angle kinematics in fruit flies. This waveform consists of constant-velocity sections, as in a triangle wave, connected by sinusoid segments at the points of velocity reversal. When this waveform is visualised in the $\ddot{x}$-$x$ space, it appears as a freeplay nonlinearity (Fig. 3A). As such, it can be generated as the solution of an additional ordinary-differential equation (ODE), controlled by a sharpness parameter $R$:

$$
\ddot{x} = h(x) = \begin{cases} -\omega^2 K(R)(x + R\hat{x}) & -\hat{x} \leq x \leq -R\hat{x} \\ 0 & -R\hat{x} \leq x \leq R\hat{x} \\ -\omega^2 K(R)(x - R\hat{x}) & R\hat{x} \leq x \leq \hat{x} \end{cases}
\tag{13}
$$

$$
K(R) = \left( \frac{2R}{\pi(1-R)} + 1 \right)^2
$$

Initial conditions: $\dot{x}(t = 0) = 0, \; x(t = 0) = \hat{x}$.

The solution, $x(t)$, to this generative free-play, nonlinear, ODE is the desired parameterised triangle-like waveform. The ODE parameters are the waveform amplitude $\hat{x}$, angular frequency $\omega$, and the sharpness parameter $0 \leq R < 1$. The latter represents the width of the freeplay region in $\ddot{x}$-$x$ space: in the time-domain, the region of constant velocity. The other region, where $\ddot{x} \propto -x$, generates sine wave segments that interpolate between these regions of constant velocity. Setting $R = 0$ generates a pure sine wave; and $R \rightarrow 1$ a pure triangle wave. Further details of the waveform formulation in the time domain are given in the appendix (A.2).

Consider then our linear PEA system (Eq. 12). We start at a prescribed linear resonant state: a simple-harmonic wave ($x_0(t), R = 0$) at the natural frequency $\omega_0$, and any amplitude, $\hat{x}_0$. By the elastic-bound condition (Eq. 8), we can confirm that this state is energy resonant ($x_0(t) \in \mathcal{X}$). If we then alter the prescribed output waveform from $x_0(t)$ to $x_R(t)$, maintaining constant $\omega = \omega_0$ and $\hat{x} = \hat{x}_0$, but altering $R$ to some $R > 0$, then the resulting wave, $x_R(t)$ no longer satisfies the PEA equality conditions (Eq. 11). Namely, the loads at the displacement extrema do not remain constant:

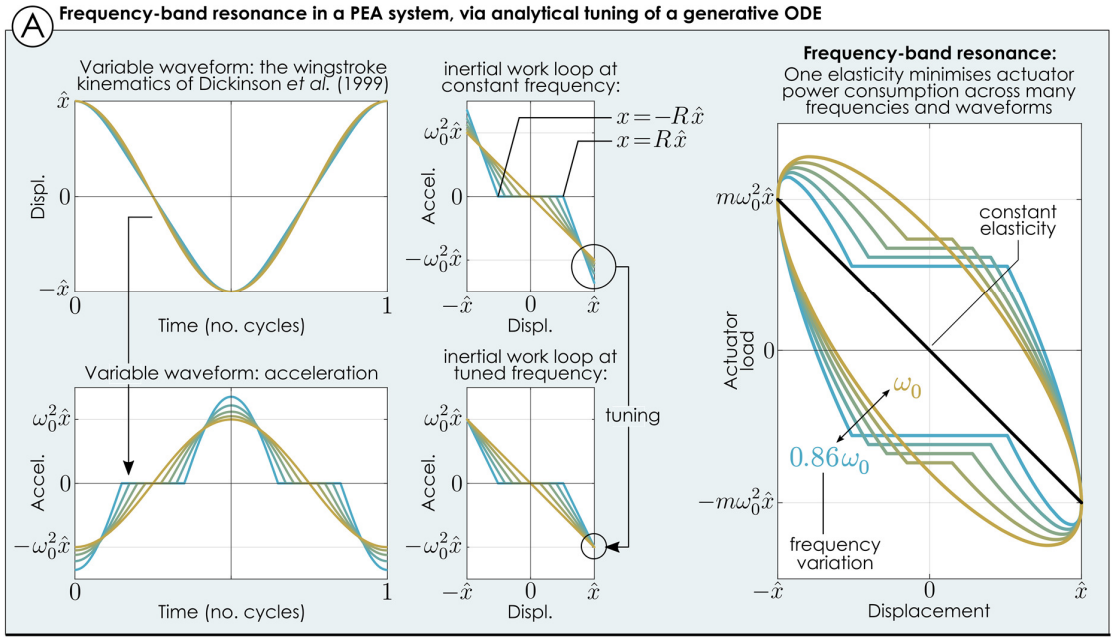

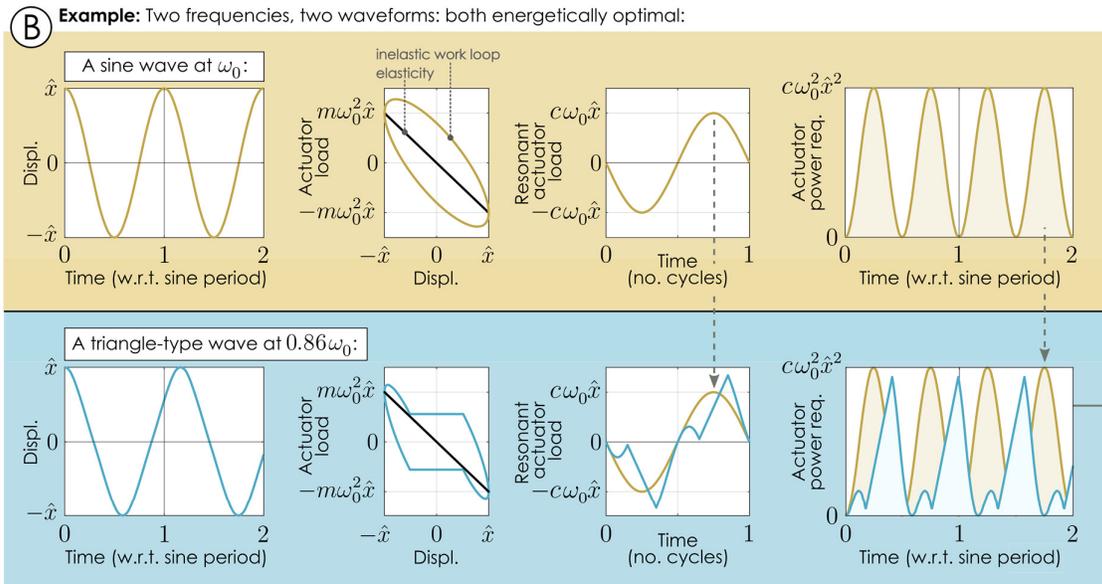

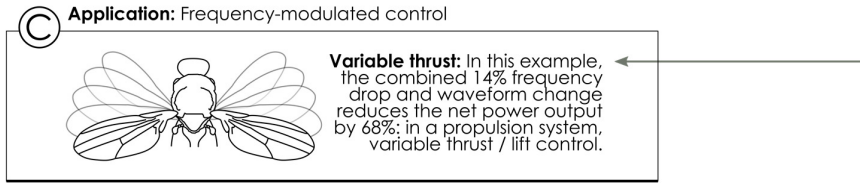

**Fig. 3: Frequency-band resonance in a PEA system, via analytical tuning of a generative ordinary-differential equation (ODE).** (**A**) Frequency-band resonance in a PEA system. For demonstration, we utilise a generalisation of the triangle-like waveform of Dickinson *et al.* [92], devised as a model of insect wing-stroke kinematics; and a linear PEA system. At a constant frequency ($\omega_0$) the peak accelerations for each waveform does not match; the wave frequency must be tuned to generate a match. When this is done, the work loops for a range of

waveforms show a common optimal linear elasticity – that is, optimal resonant power savings are available at a range of frequencies and waveforms. (**B**) For further demonstration, two specific frequency-band resonant states are detailed: a sine wave at reference frequency $\omega_0$, and a triangle-like wave at frequency $0.86\omega_0$; the extremal cases. The modulation in resonant load and power waveform for each are illustrated – note the absence of negative power in both cases. (**C**) An example application: frequency-modulated thrust or lift control in biomimetic propulsion systems, at optimal mechanical efficiency.

For $x_0(t)$: *e.g.*, $G^{\pm}(\hat{x}_0) = mh(\hat{x}_0) = -m\omega_0^2\hat{x}_0$.

For $x_R(t)$: *e.g.*, $G^{\pm}(\hat{x}_0) = mh(\hat{x}_0) = -m\omega_0^2 K(R)(1-R)\hat{x}_0 \neq -m\omega_0^2\hat{x}_0$. $\quad$ (14)

$x_R(t)$ is thus not energy resonant ($x_R(t) \notin \mathcal{X}$). To ensure at least that the PEA equality conditions (Eq. 11) are satisfied, starting again from $x_0(t)$, we alter $\omega$ and $R$ simultaneously and use our generative ODE (Eq. 13) to obtain a new waveform, $x_{R,\omega}(t)$. For this waveform, at general $R$ and $\omega$, the loads at the displacement extrema are:

$$G^{\pm}(\hat{x}_0) = mh(\hat{x}_0) = -m\omega^2 K(R)(1-R)\hat{x}_0 = -G^{\pm}(-\hat{x}_0). \quad (15)$$

Thus, so ensure these loads (Eq. 15) are equivalent to the linear resonant loads (Eq. 14), we require that, at any $R$, the frequency is determined by $\omega = \Omega(R)$, where:

$$\Omega(R) = \frac{1}{\sqrt{(1-R)K(R)}}\omega_0. \quad (16)$$

Note that $\Omega(R) \leq \omega_0, \forall R$. Eq. 16 is an analytical tuning relation, generating a continuous set of tuned waveforms, $x_{R,\Omega(R)}(t)$, of varying frequency and sharpness, and satisfying the PEA equality conditions (Eq. 11).

We then compute the range of $R$ for which $x_{R,\Omega(R)}(t) \in \mathcal{X}$: that is, $R$ for which the tuned waveforms, $x_{R,\Omega(R)}(t)$ satisfy the full PEA elastic-bound inequality conditions, and are thus energy resonant. We observe that $x_{R,\Omega(R)}(t)$ satisfying these inequality conditions, satisfy them over the range $0 \leq R \leq R^*$, for some critical $R^* \geq 0$. This range yields a frequency range $\omega^* \leq \omega \leq \omega_0$, for some critical $\omega^* = \Omega(R^*)$. It is possible to compute $R^*$ analytically, under the observation (*cf.* Fig. 3A) that the violation of the PEA inequality condition, *i.e.*, the intersection of $-F_s(x)$ and $G^{\pm}(x)$, always occurs first at $x = \pm R^*\hat{x}$. Solving for $R^*$ in $-F_s(R^*\hat{x}) = G^{\pm}(R^*\hat{x})$ yields the critical values:

$$R^* = \frac{-c^2 + c\sqrt{c^2 + 4m^2\omega_0^2}}{2m^2\omega_0^2} = -2\zeta^2 + 2\zeta\sqrt{1+\zeta^2}.$$

$$\omega^* = \Omega(R^*) = \frac{1}{\sqrt{(1-R^*)K(R^*)}}\omega_0 = \frac{\pi}{(4-\pi)\zeta + \pi\sqrt{1+\zeta^2}}\omega_0. \quad (17)$$

The tuned waveform generated by Eq. 13 is energy resonant for this PEA system over the entire range $0 \leq R \leq R^*$, $\omega = \Omega(R)$, $\omega^* \leq \omega \leq \omega_0$, and for $\hat{x} = \hat{x}_0$. The associated input load requirement is the associated $F(t)$ given in Eq. 12.

Figure 3A illustrates this process graphically. The system dimensional parameters are, $m = 1$, $c = 0.8$, $k = 1$; or, as an equivalent canonical oscillator, $\omega_0 = 1$ and damping ratio $\zeta = 0.4$. We prescribe oscillations with amplitude $\hat{x} = \hat{x}_0 = 1$. Under these conditions, the analytical results for the critical waveform parameters are $R^* \cong 0.54$ and $\omega^* \cong 0.84\omega_0$. A graphical approach confirms these values: we can easily obtain a conservative estimate for $R^*$ by incrementing $R$ over $0 \leq R < 1$ and observing when there is an intersection between the inelastic work loop, $G(t)\text{-}x(t)$, and the linear elasticity $F_s(x)$. Using this approach, we estimate $R^* \cong 0.51$ and $\omega^* \cong 0.86\omega_0$, as illustrated in Fig. 3A. Thus, we have obtained frequency-band resonance: output waveforms with $0 \leq R \leq 0.5$ are energy resonant, and span a range of frequencies, $0.86\omega_0 \leq \omega \leq \omega_0$. This range of frequencies is specific to this particular PEA system, and the choice of the triangle-type variable waveform (Eq. 13). This range is not, in fact, the widest range available for this particular system under *any* possible variable waveform. Heuristically, we may observe that a wider range of frequencies would be available if we used different waveforms generated by smoother acceleration functions in the generative ODE, for instance, $\ddot{x} \propto -x^3$. There, the curve $x^3$ describes the inertial component of the inelastic work loop ($G$-$x$), *i.e.*, the midline of $G^{\pm}(x)$. Geometrically, a midline with strong changes in gradient, *i.e.*, pronounced peaks will allow smaller changes in frequency, before $G^{\pm}(x)$ intersects the elasticity, $F_s(x)$, violating the elastic-bound condition (Fig. 3A). A smoother midline will, instead, delay condition violation, and is thus preferable.

These results are intriguing. Even in a perfectly linear system, by using relatively small changes in waveform (Fig. 3A), we can change the frequency of the system output, within a certain band, and still maintain perfect energy transfer, *i.e.*, a complete absorption of inertial power requirements. At each of these frequencies and waveforms, the fixed system elasticity is energetically optimal: it is impossible to decrease the actuator mechanical power consumption, under any metric (Section 2.2), by choosing any other elasticity. For the example PEA system and waveform, we can alter[3] the output frequency by up to 14% of the initial harmonic frequency, corresponding to a 32% modulation in the net power flow through the system, while maintaining energetic optimality. Consider an application of this principle in a biomimetic vibrational propulsion system: this frequency modulation capability could represent continuous lift or thrust control, or the operation of a propulsion system at two distinct frequencies (Figs. 3B-C). For instance, we might select two extreme frequency-band states as two modes of operation for a MEMS micro-swimmer robot: a harmonic wave at $\omega_0$, and a frequency-band resonant state at $0.86\omega_0$. One fixed elasticity would allow energetically-optimal operation at both frequencies, providing us with two propulsive modes, high-thrust and low-thrust. Indeed, the linear energy resonant states presented in this section are particularly relevant to micro-swimmer robots, for which dissipative forces may be linear (Eq. 12) [94].

We can extend this concept further, by contemplating the possibility of continuously passing a system through frequency-band resonant states, and thus, continuously modulating the system operating frequency. In an insect, or an FW-MAV, this would represent continuous lift control via frequency-modulation, at a state of continuous energetic optimality – a novel form of control, which is consistent with similar control behaviour observed in insects. Honeybees, fruit flies, and hawkmoths are observed to modulate wingbeat frequency as a mechanism of aerodynamic force control [36, 57–59]; mosquitos are observed to do so in acoustic courtship interactions [69, 95]; and many species of bee and fly show temperature-dependent wingbeat frequency variation that is though to be a form of internal temperature regulation [60, 96]. Quantitively, the linear model studied in Fig. 3, with $\zeta = 0.4$, is a broad approximation of a strongly-damped insect, e.g., a fruit fly [20]. Fruit flies have been observed to show wingbeat frequency modulation of up to 15%; a value consistent with our estimate of 14% modulation under continual energetic optimality. Frequency-band resonance may indeed provide an explanation for this biological control behaviour – an explanation which is valid for both linear and nonlinear models of the insect flight motor.

---

[3] A technical point: in this PEA system, a triangle-type waveform defined by Eq. 13 allows only a reduction in frequency relative to the initial harmonic wave. Obtaining an increase in frequency is relatively simple. We can (**i**) define the initial state as multiharmonic ($R \neq 0$) instead of single-harmonic ($R = 0$), or (**ii**) choose an alternative generative ODE, $\ddot{x} = h(x)$: *e.g.*, $h(x) \propto -\operatorname{sgn} x\, |x|^n$, $0 < n \leq 1$, which generates increases in frequency over a sine wave ($n = 1$). There is scope for a theoretical treatment of the relationship between $h(x)$ and the frequency-band resonant states it generates – we leave such a treatment to future work.

In addition, frequency-band resonance may provide insight into the nominal operating state of biological propulsion systems, such as insect flight motors. Given that insect wing stroke kinematics in steady level flight are not exactly simple-harmonic [72, 92, 93, 97–99], frequency-band resonance implies that the energy resonant state of the real flight motor may *not* be at the same frequency as the harmonic energy resonant state (i.e. $\Omega(R) \leq \omega_0$, Eq. 16). For instance, consider modelling a flight motor as a strongly-damped PEA system with linear elasticity, and with wingbeat kinematics, $x(t)$, described by a wave with nonzero sharpness parameter – e.g., $R \approx 0.6$ [92, 99]. The frequency at which these wingbeat kinematics are most efficiently generated, in terms of minimum power consumption, is lower than the flight motor natural frequency: $\Omega(R) < \omega_0$. For $R \approx 0.6$, we can estimate $\Omega(R) \approx 0.80\omega_0$. This is an approximate estimate, based on linear, and purely parallel, elasticity.

To gain a broader understanding of this effect, we can examine the critical frequency and sharpness, $\omega^*$ and $R^*$ (Eq. 17), of our linear PEA model, as a function of the structural damping ratio, $\zeta$. This puts loose bounds on the maximum possible waveform sharpness, and minimum waveform energy resonant frequency, of a simple PEA system – for instance, a model of an insect flight motor. Figure 4 shows an illustration of these bounds, tracking the critical frequency, $\omega^*(\zeta)$, over $\zeta$, alongside the three classical frequencies: the undamped natural frequency, $\omega_0$; the damped natural frequency, $\omega_d(\zeta)$; and the harmonic resonant frequency, $\omega_r(\zeta)$ [100]. Several interesting points can be noted. (**i**) The band-width of the frequency-band states increases with damping, approximately linearly. (**ii**) In the case of an insect flight motor, these effects could explain discrepancies between measured motor natural frequencies and insect wingbeat frequencies [34]. Based on the relationship between damping ratio and $Q$-factor, $\zeta \approx 1/2Q$ [20, 101], and existing estimates of the $Q$-factor for insect flight [20, 37], we would estimate $0.3 \leq \zeta \leq 0.4$ for a wide range of insect species. This would correspond to a frequency drop of 10-15% with respect to the natural frequency, which is a significant proportion of the frequency mismatch observed in [34]. And finally, (**iii**) This analysis can be repeated for more accurate models of insect flight motors, e.g., with quadratic damping, computational models of fluid-structure-interaction, or experimental similitude models. Such analyses would yield more accurate estimates of the explanatory power of frequency-band resonance in insects, and its potential capability in FW-MAVs. The principles of frequency-band resonance that we have studied in this system transfer directly to nonlinear PEA systems (Sections 3.1-3.2), for which numerical analysis techniques may be more practical, as demonstrated in Section 4.3. Applications for frequency-band resonance in nonlinear systems form a broad constellation of possibilities, which we have only just begun to observe.

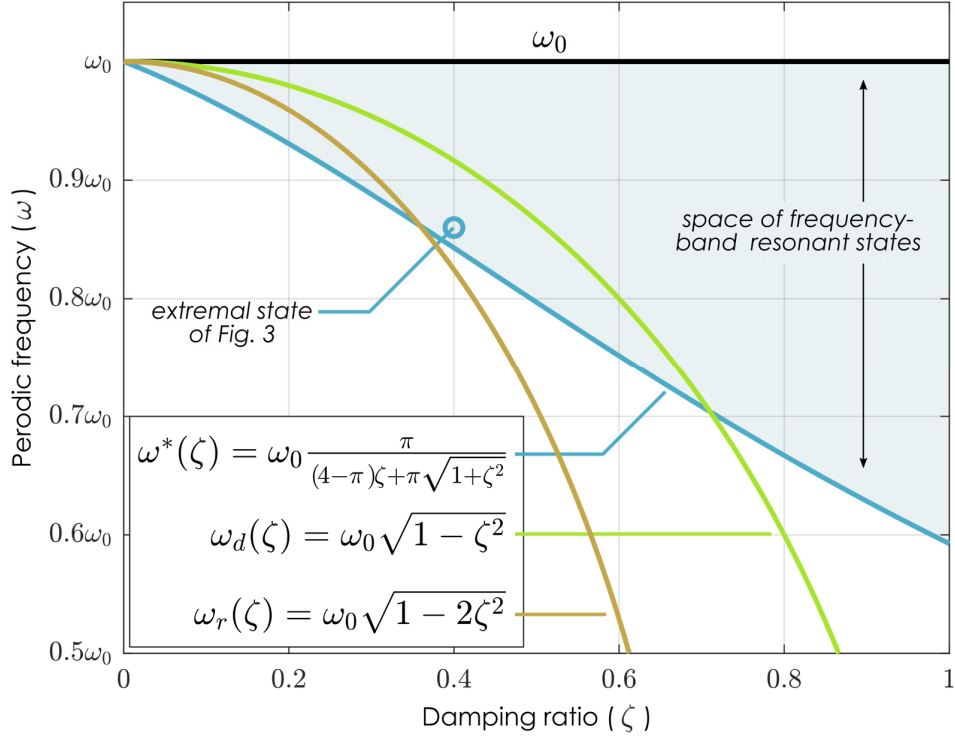

**Fig. 4: The critical frequency and three classical frequencies, as function of damping ratio.** The critical frequency, $\omega^*$, is the lower bound of the space of frequency-band resonant states (in $R$), *i.e.*, the minimum frequency for which an energy resonant state exists at some waveform. In a linear system, it is distinct from the three classical frequencies – the natural frequency, $\omega_0$; damped natural frequency, $\omega_d(\zeta)$; and harmonic resonant frequency, $\omega_r(\zeta)$ – but is of similar magnitude.

## 4.2. Offset-band resonance in a PEA system, via analytical spectral tuning

In the same system used in Section 4.1 – a general time-invariant 1DOF linear system, under PEA, with fixed inertia, dissipation, and elasticity – we find also the phenomenon of offset band-resonance. Within the set of band-type resonant states for this system, there are not only symmetric states at different frequencies (frequency-band resonance), but, in addition, symmetry-broken states: states with non-symmetric output, $x(t)$. As a test case, we focus on a form of offset-band resonance that is directly relevant to insect flight (Fig. 5): the case where one extremum of the output waveform is held fixed (constant $\min x$) and the other is modulated (variable $\max x$). This symmetry-breaking modulation corresponds to a form of insect body pitch control: symmetric modulation of only the forward-stroke position of the two wings [62, 63]. Here, we illustrate an approach for constructing these offset-band resonant states based on analytically tuning a multiharmonic waveform – a method particularly relevant to control applications. Beginning with a simple-harmonic waveform, $x(t) = \hat{x}\cos(\omega_0 t)$, at the system linear resonant frequency, $\omega_0$, and some initial amplitude, $\hat{x}$, we desire an extension into asymmetric amplitudes ($\max x \neq -\min x$), such that: **(i)** the displacement minimum is constant ($\min x = -\hat{x}$), and the displacement maximum is modulated by some $\Delta x$ ($\max x = \hat{x} + \Delta x$); and **(ii)** the PEA equality conditions (Eq. 10) are satisfied.

As a candidate for such a waveform, we select a multiharmonic wave, with fundamental period $T_0 = 2\pi/\omega_0$, and containing cosine terms of even order (to ensure symmetry about $t = T_0/2$):

$$x(t) = a_1 \cos(\omega_0 t) + a_2 \cos(2\omega_0 t) + a_4 \cos(4\omega_0 t), \tag{18}$$

The intuition behind this particular choice is that even harmonics, $\cos(n\omega t)$ with $n$ even, alter the waveform extrema asymmetrically, and therefore are well-suited to offset-band resonance. We will need one fundamental harmonic ($\omega_0$); one harmonic ($2\omega_0$) just to ensure modulation of $\max x$ independent of $\min x$; and at least one additional harmonic ($4\omega_0$) to satisfy the PEA equality condition. To compute appropriate coefficients, $a_1$, $a_2$ and $a_3$, we define conditions for the two cycle extrema at $t = 0$ and $t = T_0/2$:

Displacement extrema:  $\quad x(0) = \hat{x} + \Delta x, \qquad\qquad x(T_0/2) = -\hat{x},$
PEA equality conditions:  $\quad \ddot{x}(0) = -\omega_0^2(\hat{x} + \Delta x), \qquad \ddot{x}(T_0/2) = \omega_0^2\hat{x}.$ $\qquad$ (19)

That is, the acceleration $\ddot{x}(T_0/2)$ is independent of $\Delta x$ whereas the acceleration $\ddot{x}(0)$ scales linearly with $\Delta x$, starting from $\omega_0^2\hat{x}$ (Fig. 5). The waveform that satisfies Eqs. 18-19 is:

$$x(t) = \left(\hat{x} + \frac{1}{2}\Delta x\right)\cos(\omega_0 t) + \frac{5}{8}\Delta x \cos(2\omega_0 t) - \frac{1}{8}\Delta x \cos(4\omega_0 t), \tag{20}$$

an expression which is obtained by substituting Eq. 19 into Eq. 18 and solving for $a_1$, $a_2$ and $a_3$. Eq. 20 is variable waveform parametrised by, $\Delta x$. Under certain conditions on $\Delta x$ this waveform will simultaneously generate the appropriate waveform kinematic change ($\min x = -\hat{x}$, $\max x = \hat{x} + \Delta x$) and satisfy the PEA equality conditions.

The conditions on this solution that we must be mindful of are as follows. (**i**) In defining acceleration conditions at $t = 0$, $T_0/2$, we have assumed that these times correspond to the displacement extrema: $\max x = x(0)$ and $\min x = x(T_0/2)$. We expect that this assumption will hold true over a certain range of coefficient values ($a_1$, $a_2$ and $a_3$): in particular, it will hold true when the fundamental harmonic is dominant ($a_1 \gg a_2, a_3$). (**ii**) However, as per Section 2.1, the conditions of our PEA-system analysis specify that $x(t)$ must be composed of two monotonic half cycles: that is, $\Delta x$ must be such that $\dot{x}(t) = 0$ *only* at $t = 0$ and $t = T_0/2$. The latter is a more restrictive condition that the former: if condition (**i**) fails, then condition (**ii**) necessarily fails, but not vice-versa. The limits on the available $\Delta x$ interval – such that Eq. 19 is admissible under our analysis, and satisfied the PEA equality conditions – are thus defined by condition (**ii**). Analytical approaches to computing these limits are unwieldy: we provide a conservative numerical estimate: $-0.4 < \Delta x/\hat{x} < 0.7$ for condition (**ii**) to be satisfied. Note that this interval is specific to this particular waveform (Eq. 20), and is not a property of band-type resonance in this system generally.

Finally, the interval in $\Delta x$ that we have just defined is only the interval such that $x(t)$, as per Eq. 20, is admissible under our analysis, and satisfies the PEA equality conditions (Eq. 11). The full PEA *inequality* conditions (Eq. 8) may be satisfied only over some subinterval of this interval (possibly, nowhere). This subinterval is the set of offset-band resonant outputs. We estimate it numerically and graphically (Fig. 5) as $-0.15 < \Delta x/\hat{x} < 0.175$ (Fig. 5). That is, a 17.5% increase, and 15% decrease in the displacement at the full-cycle point ($\max x$) are available, at continuous power optimality, and at an output waveform which shows only small deviations from a simple-harmonic wave (Fig. 5).

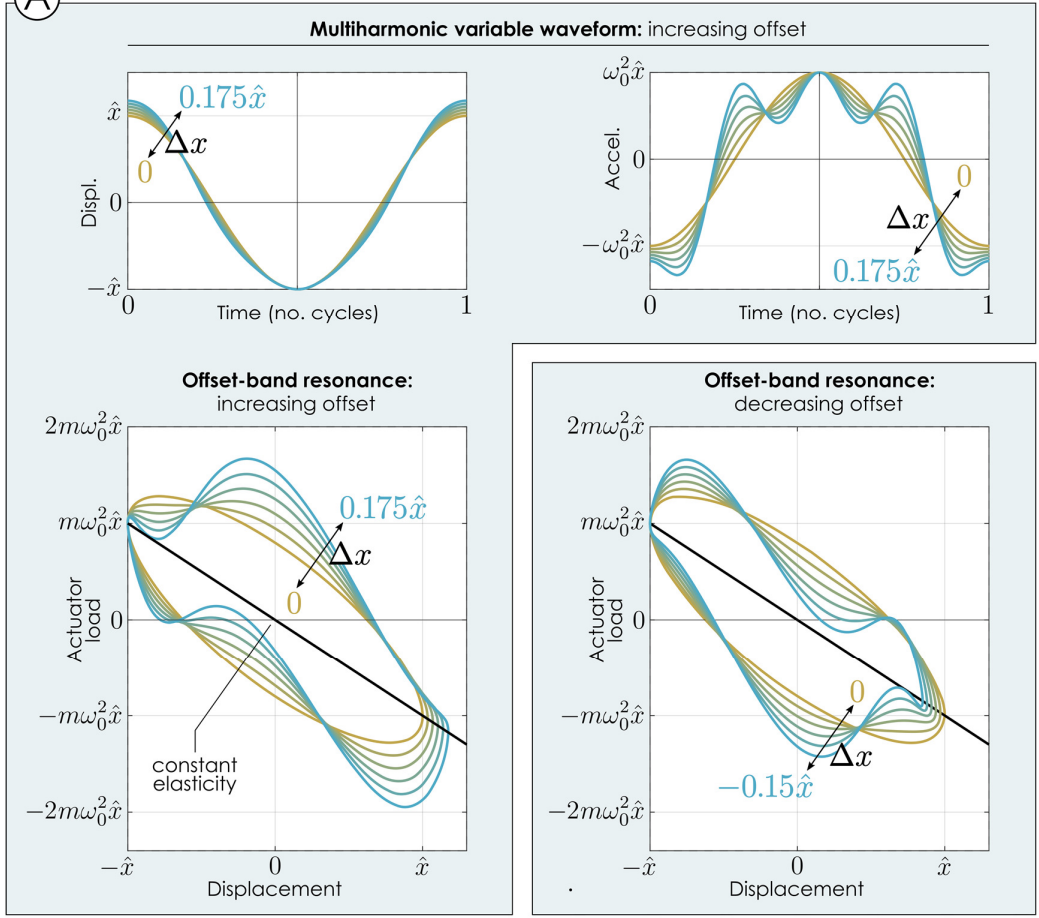

**Multiharmonic variable waveform:** increasing offset

**Offset-band resonance:** increasing offset

**Offset-band resonance:** decreasing offset

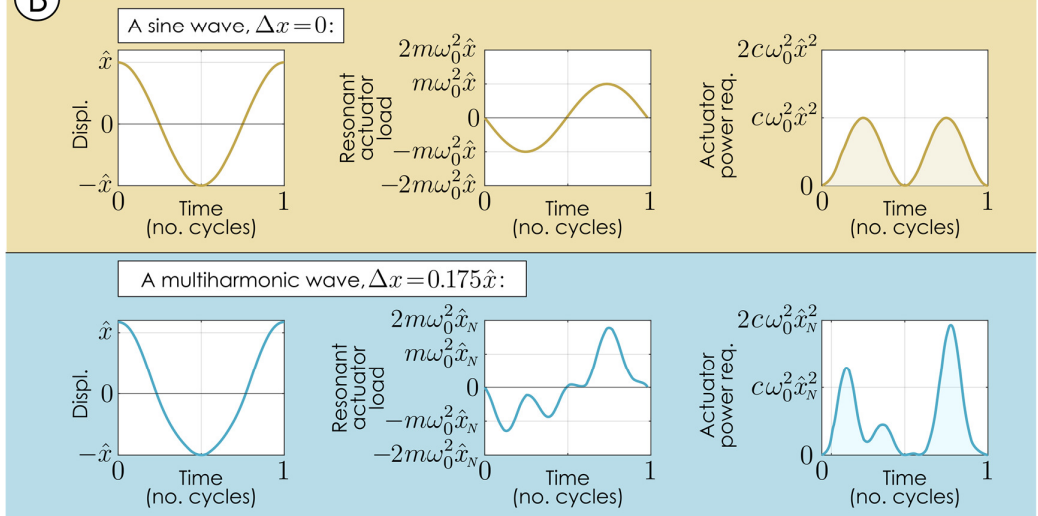

A sine wave, $\Delta x = 0$:

A multiharmonic wave, $\Delta x = 0.175\hat{x}$:

 **Application:** Vectored-thrust control in propulsion systems. Offset variation allows the alteration of the direction associated with propulsive thrust, *e.g.*, lift in an insect, or FW-MAV system.

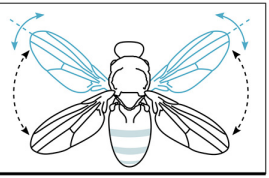

**Fig. 5: Offset-band resonance in a PEA system**. (**A**) Offset-band resonance in a PEA system. For demonstration, we utilise a multiharmonic waveform as per Eq. 18; and a linear PEA system. By default, this multiharmonic waveform satisfies the condition for acceleration-matching at the displacement extrema ($-\hat{x}$ and $\hat{x} + \Delta x$); and thus, the work loops for a range of waveforms show a common optimal linear elasticity – optimal resonant power savings are available at a range of single-point offsets ($\Delta x$), both positive and negative. (**B**) For further demonstration, two specific offset-band resonant states are detailed: a sine wave, and a multiharmonic wave at $\Delta x = 0.175\hat{x}$; the positive extremal case. The modulation in resonant load and power waveform are illustrated – note the continued absence of negative power. (**C**) An example application: vectored-thrust control in propulsion systems, *e.g.*, FW-MAVs.

In vibrational propulsion systems, such as FW-MAVs [48], micro-swimmers [90], and droplet propulsion devices [91], the symmetry-breaking character of offset-breaking resonance is of key importance. In these systems, offset-band resonance could allow resonant thrust-vectoring: position and orientation control via modulation of the distribution of lift, thrust, and other propulsive forces in space and time [62, 102]. For instance, the particular form of offset control studied in this section is identical to that used by fruit flies for pitch-axis control [62, 63]. The offset ($\Delta x$) modulation reported by Whitehead et al. [62] for fruit flies lies within the range $-0.4 < \Delta x/\hat{x} < 0.4$. Our predicted frequency-band range for a broadly representative fruit fly flight motor, with $\zeta = 0.4$, explains just under half of this reported offset modulation range – at the least, significantly alleviating the energetic cost of this control modulation behaviour. In FW-MAV systems, a sufficient characterisation of the system stability and controllability may allow the development of a similar pitch-axis continuous controller based on offset-band resonance. Indeed, in continuous-control applications, the multiharmonic approach to generating these offset-band states may have a significant advantage over the generative ODE approach (Section 4.2): the control space in actuator load is simple and easily accessible. That is, the only control parameters required are the components of the actuator load at frequencies $\omega_0$, $2\omega_0$, and $4\omega_0$. And finally, we note that a wide range of other forms of offset-band phenomena are also available using only a few harmonics: for instance, the symmetric waveform shift from $x \in [-\hat{x}, \hat{x}]$ to $x \in [-\hat{x} + \Delta x, \hat{x} + \Delta x]$. This symmetric shift may be particularly relevant to applications requiring less coupling between thrust and steering [103]: unlike the biomimetic single-point control studied in this section, the symmetric shift maintains a more constant overall power output (*e.g.*, a more constant lift force in a FW-MAV).

### 4.3. Frequency-band resonance in an SEA system, via numerical spectral tuning

As per Section 3.1, band-type resonance is also available in SEA systems. The multiharmonic approach introduced in Section 4.2 is a suitable methodology for constructing these band-type resonant states, and is demonstrated here as a numerical approach. For SEA frequency-band resonance, we select the following parametrised waveform:

$$x(t) = \hat{x}\big(1 - \alpha(\omega)\big)\cos(\omega t) + \hat{x}\alpha(\omega)\cos(3\omega t)\,, \tag{21}$$

with constant amplitude $\hat{x}$; frequency $\omega$, which will be modulated; and a tuning function $\alpha(\omega)$. The intuition behind this choice of harmonic is that odd harmonics, $\cos(n\omega t)$ with $n$ odd, alter the waveform extrema symmetrically, and are therefore well-suited for frequency-band resonance. As the magnitude of a single harmonic can be tuned to the frequency (Eq. 21), we typically require only one odd harmonic for frequency-band resonance. Given that symmetry is preserved, tuning frequency-band resonance is typically easier than offset-band resonance – to the extent that we can devise a robust numerical method for this process, as follows.

Consider a linear SEA system:

| | |
|---|---|
| System dynamics: | $D(\dot{x}, \ddot{x}) = m\ddot{x} + c\dot{x}.$ |
| Load requirement: | $F(t) = D(\dot{x}, \ddot{x}) = m\ddot{x} + c\dot{x}.$ |
| Fixed elasticity: | $F_s(u - x) = k(u - x).$ |
| Natural frequency: | $\omega_0 = \sqrt{k/m}.$ |
| Damping ratio: | $\zeta = c \,/\, 2\sqrt{km}.$ |
| Energy resonant frequency: | $\omega_e = \omega_0\sqrt{1 - 4\zeta^2}.$ |

(22)

This linear SEA system can be used as a simple model of MEMS systems under base-excitation [28]; atomic force microscopes [29, 30]; certain insect flight motors [20]; and prosthetic and robotic leg structures [54, 55]. We select parameters as per the PEA system in Section 4.1: $m = 1$, $c = 0.8$, $k = 1$; *i.e.*, natural frequency, $\omega_0 = 1$ and damping ratio $\zeta = 0.4$. We note that, in this system, the frequency at which a pure sinusoidal wave is energy resonant is $\omega_e$, rather than $\omega_0$ – these frequencies coincide in a linear PEA system, but not in a linear SEA system [20]. As per the spectral shaping process (Section 3.2) we seek a tuning function, $\alpha(\omega)$, such that the system satisfies the SEA equality conditions (Eq. 11), at least over a relevant range of $\omega$. We will characterise this function numerically. At a specified $\omega$, we formulate an algorithm for computing $\alpha(\omega)$ – in this linear SEA system, as well as other nonlinear SEA systems. Such an algorithm can be constructed in a number of ways, but must in general contain two components: (i) a method for estimating the SEA work loop gradient $X'^{\pm}(F = 0)$ (Eq. 11), for some given value of $\alpha$ and $\omega$, and thus, for estimating the error in the SEA equality condition at this $\alpha$ and $\omega$; and (ii) a method for solving for a value, $\alpha(\omega)$, such that this error is minimised. For the first component, our particular implementation uses the chain-rule for evaluating $X'$:

$$X'\big(F(t)\big) = \frac{d}{dF} X\big(F(t)\big) = \frac{d}{dF} x(t) = \frac{d}{dt} x(t) \left/ \frac{d}{dt} F(t) = \frac{\dot{x}(t)}{\dot{F}(t)}.\right.$$

(23)

That is, given $\alpha$ and $\omega$, we begin with, $\dot{x}(t)$ and $\dot{F}(t)$, functions of time, as evaluated from the waveform definition (Eq. 21) and system dynamics (Eq. 22). We can translate these functions of time to evaluations at force, and particularly, at $F = 0$, for $X'^{\pm}(0)$. To do this, we must compute the critical time-values $t^* : F(t^*) = 0$. This can be done using numerical root-finding: preferably, through a globally-convergent method, as there will typically be at least two roots of interest, associated with $X'^+$ and $X'^-$, respectively. With this $t^*$, we can compute $X'^{\pm}(0)$ via Eq. 23, and estimate the error in the SEA equality condition, Eq. 11: $E(\alpha) = X'^{\pm}(0) + (F_s^{-1})'(0)$, where $(F_s^{-1})'(0)$ is computed from the known system elasticity. To solve then for the value of $\alpha(\omega)$, as per our second component, we use a root-finding method based on interpolation to compute the zeros of the error, namely $\alpha^*$ such that $E(\alpha^*) = 0$ over a relevant interval of the value $\alpha(\omega)$. Typically, there is only one zero, and we take $\alpha(\omega) = \alpha^*$.

This numerical method is given in Algorithm 1 below. Its results, for the linear system of Eq. 22, are illustrated in Fig. 6: the system's inelastic work loops at different frequencies, tuned according to $\alpha(\omega)$, and the load, displacement, and power waveforms. The tuning process ensures that these work loops represent a constant output displacement amplitude, but differing input/output displacement waveforms. As such, these work loops represent also differing force amplitudes and waveforms. These waveforms are illustrated in Fig. 6A. In the limit state, $\omega =$ $0.655\omega_0$, the displacement waveform, $x(t)$ is composed of monotonic half cycles, but the force waveform, $F(t)$ is on the cusp of breaking this monotonicity condition. We can confirm that all these states are indeed energy resonant in two ways. First, we can confirm that the elastic-bound conditions (Eq. 6) are satisfied, by observing $X'(F)$. Second, we can observe the power requirement waveform, $P = F\dot{u}$, itself (Fig. 6B). In the inelastic case, in which the system's fixed elasticity is taken as absent, negative power persists in the system. In the elastic case, in which the system's fixed elasticity is present, this negative power is eliminated: these states are energy resonant.

Algorithm 1 is also directly applicable to nonlinear systems. Fig. 7 shows an application of this algorithm to a SEA system with quadratic damping: $D(\dot{x}, \ddot{x}) = m\ddot{x} + c\dot{x}|\dot{x}|$, with $c/m = 0.8$. Quadratic damping of this form could represent, for example, a higher-fidelity model of an insect flight motor or FW-MAV [37]. In this system, frequency-band resonant states exist as low as $\omega = 0.80\omega_e$: at least 20% frequency-reduction with respect to the harmonic energy resonant frequency (at $\alpha = 0$). Note that, in both these linear and nonlinear examples, Algorithm 1 appears to continue to generate band-type resonant states even beyond the monotonicity condition: for $F(t)$ that is *not* composed of two monotonic half cycles (Section 2.1). In the linear analysis (Fig. 6), although the monotonicity condition was violated for $\omega < 0.655\omega_e$, the algorithm produced energy resonant states for lower values of $\omega$ (not shown). In the nonlinear analysis (Fig. 7), examining the waveforms for $\omega = 0.85\omega_e$ and $0.80\omega_e$ indicates that they violate the monotonicity condition but are still energy-resonant, as they are free from negative power.

The significance of the SEA band-type resonant states studied in this section is thus twofold. In a theoretical context, these states demonstrate the breadth of band-type resonance as a dynamical-systems phenomenon. They illustrate, for instance, that band-type resonance is not simply a property of parallel-elastic (PEA) systems, but a much broader phenomenon. We can countenance the existence of band-type resonance in hybrid systems [20], and systems with multiple actuators [68]. In addition, the existence of band-type resonant states outside of the monotonicity conditions that we have applied in our analysis points to further extensions to our theoretical framework. In a practical context, these SEA band-type resonant states are directly relevant to several practical control problems: wingbeat frequency modulation in insects and FW-MAVs showing a series-elastic character [20]; efficient modulation of walking speed in series-elastic legged robots [53]; and frequency control of atomic force microscopes [29, 30]. Our numerical method for locating band-type resonant states is particularly relevant in a practical context, as it is applicable to a wide range of systems – and could be a key component in closed-loop control strategies for energetically-optimal frequency modulation in realistic, uncertain systems.

**Algorithm 1:** A numerical algorithm, based on interpolation, for spectral shaping in SEA systems: valid for $\alpha(\omega)$ (Eq. 21), or other single-parameter variable waveforms. The algorithm input is a frequency, $\omega$, at which a candidate energy resonant state (*i.e.*, a state satisfying the SEA equality conditions, Eq. 11) is desired; its output is an estimation of the appropriate tuning function value $\alpha(\omega)$.

0  **Input:** $\omega$ at which an energy resonant state is desired.

1  Define basic functions $x(t, \alpha)$, $\dot{x}(t, \alpha)$, $F(t, \alpha)$, and $\dot{F}(t, \alpha)$, as functions of time $t$ and tuning parameter $\alpha$, from waveform definition and system dynamics (Eqs. 21-22)

2  Define gradient function: $\partial x / \partial F \, (t, \alpha) = \dot{x}(t, \alpha) / \dot{F}(t, \alpha)$ (Eq. 23)

3  Define a time-series vector $\mathbf{t} \in [0, T]$, $T = 2\pi/\omega$, and an $\alpha$-vector, $\boldsymbol{\alpha}$, over some suitable interval of $\alpha$.

4  Define and compute system critical compliance $(F_s^{-1})'(F = 0)$, from known system elasticity.

5  **For** $\alpha_i \in \boldsymbol{\alpha}$:

6      Find the values of $t^*$ by solving $F(t_{ij}^*, \alpha_i) = 0$. We do this by evaluating $\mathbf{F}_i = F(\mathbf{t}, \alpha_i)$, constructing an interpolant $\tilde{F}(t, \alpha_i)$ from $\mathbf{F}_i$ and $\boldsymbol{\alpha}$, and evaluating $t_{ij}^* : \tilde{F}(t_{ij}^*, \alpha_k) = 0$.

7      **Check:** that there are only two $t_{ij}^*$ ($j \in \{1,2\}$). If not, the value of $X_i'(0)$ is undefined and we proceed to the next $\alpha_i$

8      Evaluate the gradient $\partial x / \partial F$ at $t_{ij}^*$: $\{X_{ij}'(0)\} = \{\partial x / \partial F \, (t_{ij}^*, \alpha_i)\}$, $j \in \{1,2\}$. This step uses the result of line 2 according to Eq. 23.

9      **Check:** that $X_{i,1}'(0) = X_{i,2}'(0)$, and if so, define this common value as $X_i'(0)$. If not, return $X_i'(0)$ as undefined.

10  **End** loop, return series $X_i'(0)$ as vector, $\mathbf{X}'(0)$.

11  Compute error $\mathbf{E} = \mathbf{X}'(0) + (F_s^{-1})'(0)$.

12  Construct an interpolant $\tilde{E}(\alpha)$, from $\mathbf{E}$ and $\boldsymbol{\alpha}$, and evaluate all $\alpha_k^* : \tilde{E}(\alpha_k^*) = 0$.

13  **Output:** all $\alpha_k^*$ are tuning parameters such that the waveform $x(t, \alpha)$ satisfies the SEA equality conditions at the specified $\omega$

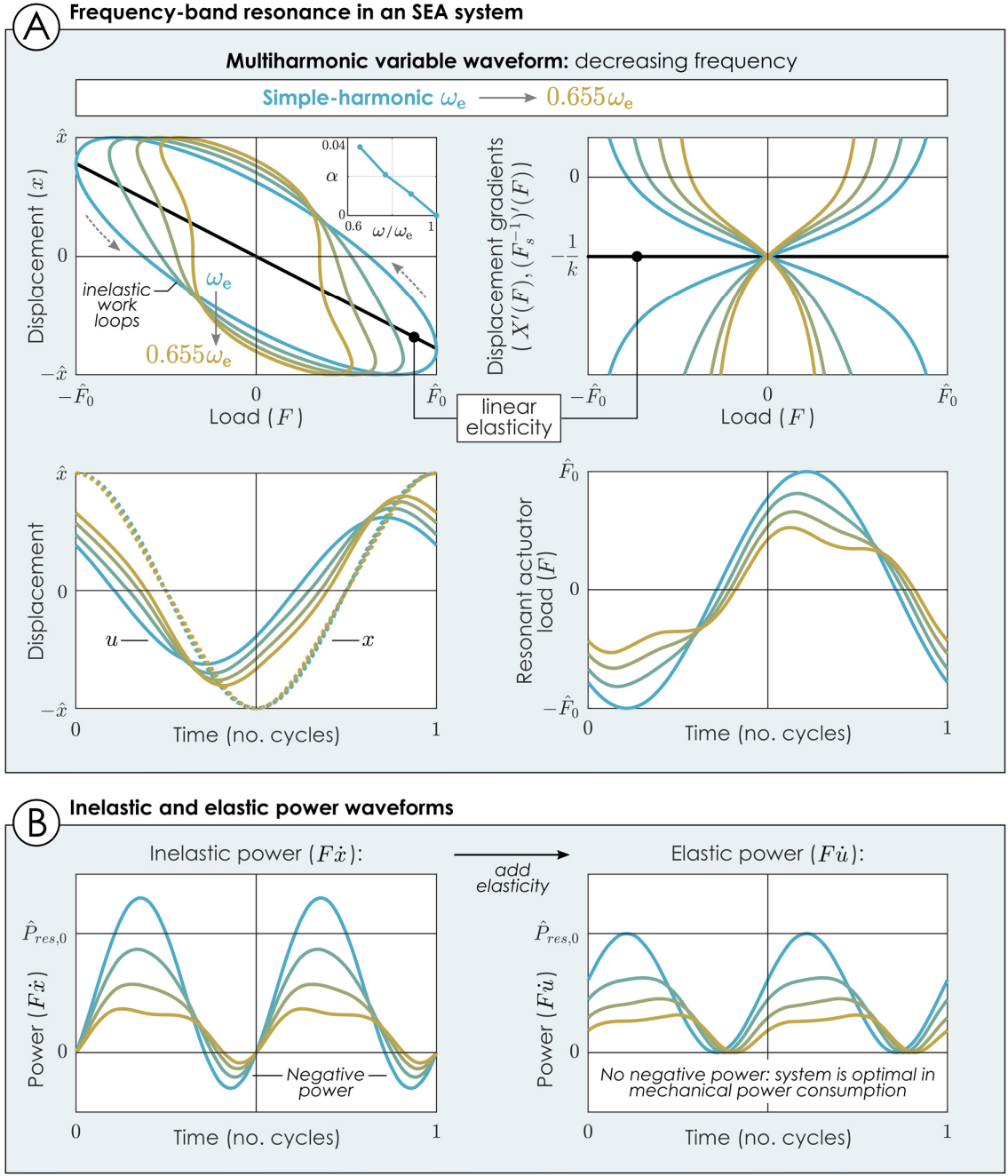

**Fig. 6: Frequency-band resonance in a linear SEA system.** (**A**) Frequency-band resonant states in a linear SEA system, computed via Algorithm 1. The SEA elastic-bound conditions (Eq. 9, Fig. 2-3) can be seen to be satisfied across the space of frequency-band resonant states. Scaling parameters are the compliance, $1/k$, and the initial simple-harmonic load amplitude, given by $\hat{F}_0^2 = \omega_e^2 \hat{x}^2 (m^2 \omega_e^2 + c^2)$. (**B**) The optimality of these states can additionally be seen in the system's power waveforms: over the space of frequency-band resonant states, the system's elastic power waveform ($F\dot{u}$) is free from negative power. The scaling parameter is the initial net power $\hat{P}_{res,0} = c\hat{x}^2 \omega_e^2$.

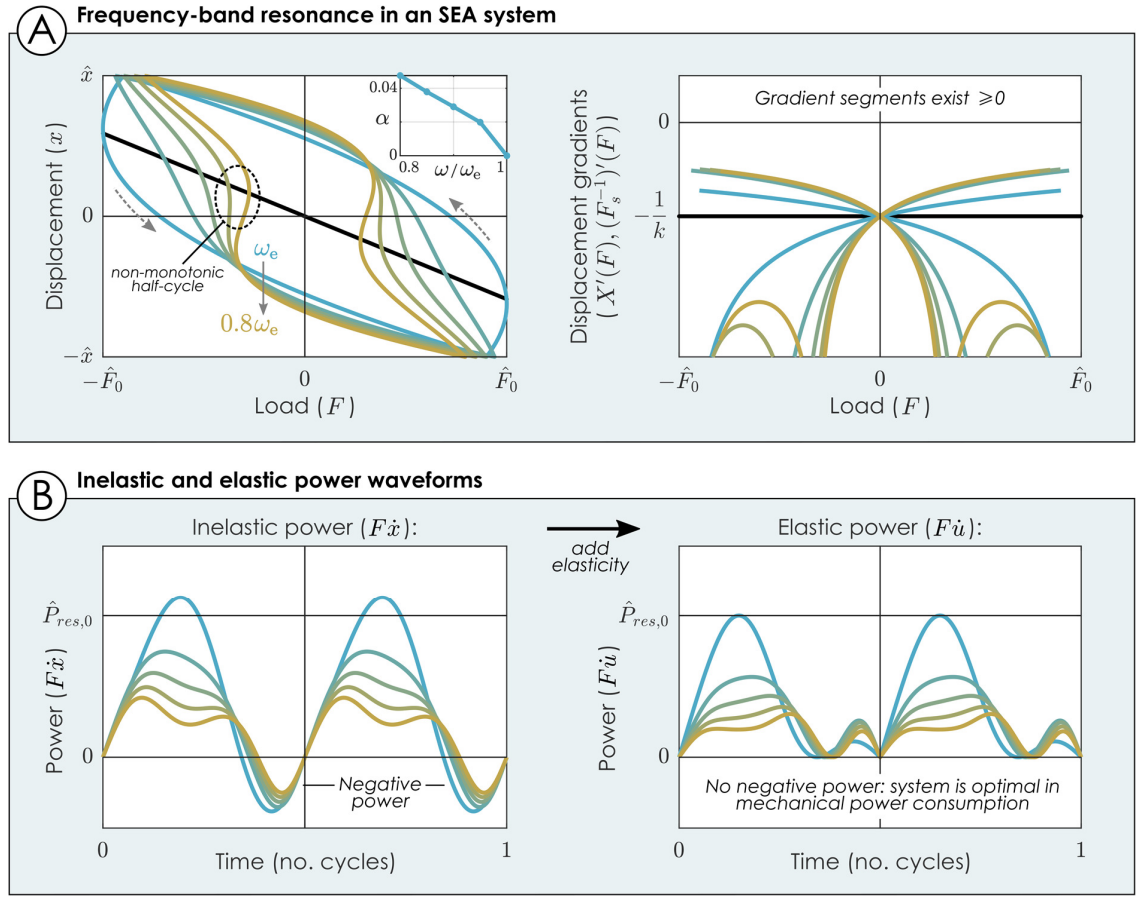

**Fig. 7: Frequency-band resonance in a nonlinear SEA system**. (**A**) Frequency-band resonant states in a quadratically-damped nonlinear SEA system, computed via Algorithm 1. Note that these states do not satisfy the monotonicity condition in $F(t)$ for $\omega = 0.85\omega_e$ and $0.80\omega_e$, as indicated. Scaling parameters are the compliance, $1/k$, and the initial simple-harmonic load amplitude, $\hat{F}_0$. (**B**) The optimality of these states, even those which do not satisfy the monotonicity condition in $F(t)$, can be seen in the system's power waveforms. Over the space of frequency-band resonant states, the system's elastic power waveform ($F\dot{u}$) is free from negative power. The scaling parameter is the initial net power $\hat{P}_{res,0}$.

## 5. Discussion

### 5.1. Implications for biolocomotion systems

The principles of frequency-band and offset-band resonance have significant practical implications for biolocomotion systems. These principles demonstrate, in a fundamental way, that frequency and offset modulation, as well as other forms of waveform modulation, are available at a state of continual energetic optimality – a state of energy resonance at fixed structural elasticity. In insects, for instance, the phenomenon of frequency-band resonance indicates that insects are able to achieve resonant oscillations of the wing-motor system not only at a single frequency, but over a range of frequencies, providing the insect can tolerate slight alterations in the wing-stroke kinematics and/or muscular load waveform as a necessary cost. A range of insect species are observed to use wingbeat frequency modulation for a range of purposes: for aerodynamic force control [36, 57–59]; for acoustic courtship interactions [69, 95]; and for temperature regulation [60, 96]. This modulation has conventionally been explained as either the result of possible real-time resonant-frequency control mechanisms within the thorax [36, 38, 65–67, 104], or in terms of the trade-off involved in deviating from resonance [36]. In contrast, our results (Section 4.1) suggest that wingbeat frequency modulation need not incur significant energetic costs – wingbeat frequency modulation via frequency-band resonance can be achieved without leaving a state of optimality in mechanical power consumption. By the same token, offset-band resonance provides explanations for mechanisms of body pitch control in insects such as fruit flies. This form of pitch-axis control involves breaking the symmetry of the resonant state: independent modulation of the wing stroke angle extrema ($\max x$ and $\min x$, as per Section 4.2). Offset-band resonance can ensure that this control is achieved at a continual state of energy resonance. Indeed, ensuring the energetic optimality of this control is made more urgent by the fact that the fruit fly free-flight pitch dynamics are unstable [63, 102], necessitating that pitch-axis control should be continually active.

Band-type resonance also elucidates a more fundamental effect in biolocomotion, and in flapping-wing flight: it illustrates how different input/output waveforms have different energy resonant frequencies. Insect wing stroke waveforms are typically not simple-harmonic, but instead can show a tringle-wave character [92]. As illustrated in Section 4.1, triangle-type waves can have significantly lower energy resonant frequencies than simple-harmonic waveforms – and this difference is of sufficient magnitude to explain resonant-frequency discrepancies observed in the literature [34]. Together, band-type resonance principles elucidate several aspects of insect flight behaviour. They apply also to the design of FW-MAVs, and other biomimetic systems: for instance, micro-swimmers driven by external acoustic or magnetic fields [49, 50], and prosthetic and robotic leg structures [54, 55].. There is significant scope for more specific studies of the role of band-type resonance in biolocomotion, as well as for further theoretical development in this area – in particular, further study of the more complex dynamical systems (*e.g.*, hybrid PEA-SEA [20], and multiple-input [68] systems) that arise in many biological locomotive structures.

## 5.2. Optimality in composite cost metrics

Band-type resonant states are energy resonant: they show optimal mechanical power consumption with respect to elasticity. In a system with some fixed elasticity, if a particular kinematic output is band-type resonant ($x_i \in \mathcal{X}$, Section 3.1), then it is not possible to choose a different elasticity such that this kinematic output can be generated at a lower overall mechanical power consumption. However, in band-type resonant states it is not only the overall mechanical power consumption which is optimal (with respect to elasticity). The principles of invariance (Section 2.3) imply that these band-type resonant states are also optimal in any invariant metrics associated with the system. For instance, in PEA systems, the absolute force metric ($\overline{P}_{|F|}$, Eq. 5) is invariant across the space of energy-resonant elasticities, if the kinematic output waveform satisfies the symmetry condition $x(t) = x(T - t), \forall t$ (Section 2.3). Therefore, if the PEA band-type resonant states satisfy this symmetry condition (*e.g.*, as in the frequency-band resonance studied in Section 4.1) then these states are optimal also in $\overline{P}_{|F|}$. It is impossible to choose a different elasticity such that $\overline{P}_{|F|}$ is further minimised. In SEA systems, any functionals of solely $F(t)$, including the force-squared metric ($\overline{P}_{F^2}$), and peak load ($\hat{F}$) are invariant across the entire space of possible elasticities. Any SEA band-type resonant state is optimal (with respect to elasticity) in any of these functionals of $F(t)$: it is impossible to choose a different elasticity such that, *e.g.*, the peak load is further minimised.

In this way, the invariance principles of Section 2.3 lead to powerful optimality principles for band-type resonant states – defining a suite of composite cost metrics which are guaranteed to be optimal with respect to elasticity, at any band-type resonant output ($x_i \in \mathcal{X}$). These optimality principles have particular practical importance in the context of choices between different frequency and offset modulation strategies. As noted in section 5.1, in insects, the action of sets of auxiliary muscles has been proposed as a mechanism for real-time tuning thoracic elasticity – potentially allowing wingbeat frequency modulation while maintaining a conventional simple-harmonic resonant state [36, 38, 65–67, 104]. Real-time elasticity tuning strategies come at additional cost in energy, actuator complexity, and control architecture, but can lead to more optimal modulated states than band-type resonant control. At any modulated state, one can (hypothetically) choose a tuned elasticity to minimise any desired cost metric. In this context, the composite-metric optimality principles for band-type resonant states help elucidate which metrics real-time elasticity tuning can further minimise, and which metrics it cannot. Over the space of band-type resonant states, invariant metrics (*e.g.*, peak force in SEA systems) cannot be minimised further by real-time elasticity tuning – hence, there is little motivation for such tuning. Metrics that are not invariant (*e.g.*, peak force in PEA systems), on the other hand, can be minimised further – it is thus possible that the implementation costs of real-time elasticity tuning could be outweighed by this additional metric minimisation.

The topic of composite-metric optimality has significant practical implications for biolocomotion. Consider two versions of a biolocomotive system (*e.g.*, a flapping-wing system): one driven by biological muscles, and one driven by an electromechanical actuator. The isometric power consumption of these two actuators – *i.e.*, the power consumption when the actuator is exerting a force, but not moving – is different. In the biological muscle, the isometric power consumption can be measured by the absolute force metric, $\overline{P}_{|F|}$, among other metrics [74, 80–82]. In the electromechanical actuator, the isometric power consumption is measured by the resistive losses; the force-squared metric, $\overline{P}_{F^2}$ [88, 89]. If this biolocomotive system is driven in a PEA architecture, then it can benefit from optimality in $\overline{P}_{|F|}$ (an invariant metric), but not in $\overline{P}_{F^2}$ (a non-invariant metric). We assume, implicitly, that these isometric power losses will be also present when the system is not static. If this is the case, the biological system will be energetically-optimal in isometric power consumption, but the electromechanical system will not. This is a crucial distinction: if, for instance, wingbeat frequency modulation strategies employed by insects take advantage of invariance-based optimality in muscular power consumption, then these same modulation strategies translated to electromechanically-driven FW-MAVs may no longer be optimal. In this way, subtle differences in the behaviour of biological and electromechanical actuators could lead to significant degradations in optimality when biological design principles are translated directly to biomimetic robots – illustrating the importance of considering these invariance principles in biolocomotive systems design and control.

## 5.3. Band-type resonance as optimal control

The band-type resonant control proposed in Section 5.1 appears to have significant connections to other areas of control theory. Firstly, we conjecture that there is a link between band-type resonance and optimal control theory. Band-type resonant states show certain optimality properties with respect to mechanical power: a cost metric sometimes expressible as a quadratic form. That is, if we represent a linear PEA or SEA system as the first-order controlled system:

$$\dot{\mathbf{x}} = A\mathbf{x} + B\mathbf{u}, \tag{24}$$

where we have $\mathbf{u} = F$ and $\mathbf{x} = [x, \dot{x}]^T$, then cost functionals based on net and absolute mechanical power, $J_{(a)}$ and $J_{(b)}$, respectively, can be defined as:

$$
\begin{aligned}
J_{(a)} &= \int_{t_0}^{t_h} \mathbf{u}^T \mathbf{Q}\, \mathbf{x}\, dt = \int_{t_0}^{t_h} F\dot{x}\, dt\,, \\
J_{(b)} &= \int_{t_0}^{t_h} |\mathbf{u}^T \mathbf{Q}\, \mathbf{x}|\, dt = \int_{t_0}^{t_h} \mathbf{u}^T [0, \mathrm{sgn}(\mathbf{u}^T \mathbf{Q}\, \mathbf{x})]\mathbf{x}\, dt = \int_{t_0}^{t_h} |F\dot{x}|\, dt\,,
\end{aligned}
\tag{25}
$$

where $[t_0, t_h]$ is the control interval and the quadratic cost matrix is $\mathbf{Q} = [0,1]$. Band-type resonant states necessarily ensure that $J_{(b)}/J_{(a)} = 1$, and the problem of locating these states can be seen as an optimal control problem to maximise $J_{(b)}/J_{(a)}$. In our analysis, we typically look for outputs, $x(t)$, that ensure energy-resonant; whereas in an optimal control context, one would look for inputs, $F(t)$; but the two approaches are closely related. For instance, consider the problem of finding any periodic $F(t)$ such that $J_{(b)}/J_{(a)}$ takes its global maximum value ($J_{(b)}/J_{(a)} = 1$). The set of solutions to this problem are contained within the system's set of band-type resonant states: they are the forcing functions, $F_i(t)$, encoded within $\mathcal{S} = \{\{x_i(t), u_i(t), F_i(t)\}\}_i$ (Section 3.1). Or, consider the more specific problem of finding a periodic $F(t)$, with specified periodic frequency $\omega$, such that $J_{(b)}/J_{(a)} = 1$. The set of solutions to this problem, if any exist, can be found within the system's set of frequency-band resonant states at $\omega$ – particular examples of which are given in Section 5.1 and 5.3, for linear PEA and SEA systems.

In our analysis, to define, *e.g.*, the frequency-band resonant states in Section 5.1 and 5.3, we assumed that the exact dynamics of the system were known. This contrasts with a more realistic control-system context, in which disturbances and unmodelled dynamics are likely to be present. To address the problem of finding band-type resonant states in realistic, uncertain systems, techniques from optimal control may be applicable. In particular, we observe that band-type resonant states resemble the sliding manifolds used in sliding-mode control [105, 106]. Sliding-mode control involves the use of discontinuous control laws to constrain a system to a favourable state-space manifold. An analogy is of a marble rolling along a narrow crack [107]: the crack walls (discontinuous control laws) constrain the system (the marble) to the sliding manifold (the crack). Given that an individual band-type resonant state can be represented as a steady-state phase portrait, *i.e.*, a state-space manifold, we propose that the techniques of sliding mode control design may be applicable to the design of controllers to constrain the system to the band-type resonant state. This is an interesting topic for future research – with relevance both for the design of controlled biomimetic propulsion systems, and the interpretation of control strategies used by biological organisms.

## 7. Conclusions

In this work, we have characterised band-type resonance: a phenomenon which represents a fundamental advance in the study of linear and nonlinear oscillators. Band-type resonance challenges several common assumptions about the nature of resonance, and the nature of energetic optimality in oscillating systems. For instance, in both linear and nonlinear systems, the state of energetic optimality is not tied to a single frequency – *e.g.*, the linear resonant frequency – but can exist in a band of frequencies. Nor is the state of energetic optimality tied to a symmetric kinematic waveform, but can be available in a band of offset, asymmetric waveforms. Even the simple result that energetically-optimal frequency modulation is available in linear PEA and SEA systems (Sections 4.1, 5.2) is a counterexample to the common supposition that deviation from the linear resonant frequency necessarily involves a loss of efficiency. Band-type resonance provides avenues for radically novel forms of resonant state control in a wide range of forced oscillators – including in biolocomotion systems, which we have considered in detail. Band-type resonance enables forms of optimal frequency (thrust / lift) and offset (steering) control in FW-MAVs and vibrational micro-swimmers that have not previously been reported. It provides alternative explanations for a range of counterintuitive control behaviours observed in insects, including in-flight wingbeat frequency and offset modulation. And it suggests that, in some situations, biological principles of optimal design might not directly translate to mechanical and electromechanical principles of optimal design. Band-type resonance has implications for biolocomotion systems, and forced oscillators wherever they may be found.

**Declarations:**

**Acknowledgements**   This work was supported, by the Israel Science Foundation grant No. 1851/17 and by the Israel Ministry of Science and Technology grant No. 3-17400.

**Data availability**     No datasets were generated or utilised in this study.

**Conflict of interest**   The authors declare that they have no conflict of interest.

# Appendix

## A.1 Closed-form expressions for PEA and SEA work loops

In Section 2.1, we defined general work-loop profiles: $G^{\pm}(x)$ and $F^{\pm}(x)$ for PEA systems; and $X^{\pm}(F)$ and $U^{\pm}(F)$ for SEA systems. These constructs can be defined analytically for a range of different systems. For a linear PEA system, with $D(x, \dot{x}, \ddot{x}, \dots) = m\ddot{x} + c\dot{x}$, $F_s(x) = kx$, and undergoing simple-harmonic oscillation according to $x(t) = \hat{x}\cos(\omega t)$, we have [25]:

$$
\begin{aligned}
G^{\pm}(x) &= -m\omega^2 x \pm c\omega\sqrt{\hat{x}^2 - x^2}, \\
F^{\pm}(x) &= kx - m\omega^2 x \pm c\omega\sqrt{\hat{x}^2 - x^2}.
\end{aligned} \tag{A.1.1}
$$

And for a quadratically-damped PEA system, with $D(x, \dot{x}, \ddot{x}, \dots) = m\ddot{x} + \mathrm{sgn}(\dot{x})\, c\dot{x}^2$, undergoing the same simple-harmonic motion, we have [25]:

$$
\begin{aligned}
G^{\pm}(x) &= -m\omega^2 x \pm c\omega^2(x^2 - \hat{x}^2), \\
F^{\pm}(x) &= kx - m\omega^2 x \pm c\omega^2(x^2 - \hat{x}^2).
\end{aligned} \tag{A.1.2}
$$

These functions are defined for $x \in [-\hat{x}, \hat{x}]$. For a linear SEA system, with $D(x, \dot{x}, \ddot{x}, \dots) = m\ddot{x} + c\dot{x}$, $F_s(x) = kx$, and undergoing simple-harmonic oscillation according to $x(t) = \hat{x}\cos(\omega t)$, the construction is more complex. For $X^{\pm}(F)$ and $U^{\pm}(F)$ we have:

$$X^{\pm}(F) = \frac{1}{m^2\omega^2 + c^2}\left(-mF \pm \frac{c}{\omega}\sqrt{\hat{x}^2\omega^2(m^2\omega^2 + c^2) - F^2}\right),$$

$$U^{\pm}(F) = \frac{F}{k} + \frac{1}{m^2\omega^2 + c^2}\left(-mF \pm \frac{c}{\omega}\sqrt{\hat{x}^2\omega^2(m^2\omega^2 + c^2) - F^2}\right), \tag{A.1.3}$$

where these functions are defined over the force range $F \in \left[-\hat{F}, \hat{F}\right]$, with $\hat{F} = \omega\hat{x}\sqrt{m^2\omega^2 + c^2}$. To compute the elastic-bound conditions, Eq. 6, one further profile is required: the gradients $X'^{\pm}(F)$. These gradients are given by:

$$X'^{\pm}(F) = \frac{1}{m^2\omega^2 + c^2}\left(-m \mp \frac{cF}{\omega\sqrt{\hat{x}^2\omega^2(m^2\omega^2 + c^2) - F^2}}\right). \tag{A.1.4}$$

This completes the definition of work-loop profiles for these example systems.

### A.2. Generalised triangle-wave function

To analyse frequency-band resonance in Section 4.1, we utilised a generalised triangle-like wave, allowing waveform alteration with associated frequency and offset modulation. This waveform was described via a freeplay-nonlinear time-invariant ODE (Eq. 11). The time-domain solution to this ODE is:

$$\ddot{x}(t) = \begin{cases} -A(\delta)\dfrac{\hat{x}}{T^2}\cos\left(\dfrac{\pi t}{2\delta T}\right) & 0 \leq t \leq \delta T \\[2mm] A(\delta)\dfrac{\hat{x}}{T^2}\cos\left(\dfrac{\pi\left(t - \frac{1}{2}T\right)}{2\delta T}\right) & \left(\frac{1}{2} - \delta\right)T \leq t \leq \left(\frac{1}{2} + \delta\right)T \\[2mm] -A(\delta)\dfrac{\hat{x}}{T^2}\cos\left(\dfrac{\pi(t - T)}{2\delta T}\right) & T(1 - \delta) \leq t \leq T \\[2mm] 0 & \text{o.w.} \end{cases} \tag{A.2.1}$$

$$A(\delta) = \frac{2\pi^2}{(8 - 4\pi)\delta^2 + \pi\delta}, \qquad \delta = \delta(R) = \frac{1}{4}\frac{\pi(1 - R)}{(2 - \pi)R + \pi},$$

with $\dot{x}(t)$ and $x(t)$ as:

$$\dot{x}(t) = \int_0^T \ddot{x}\,dt, \qquad x(t) = \int_0^T \dot{x}\,dt + \hat{x}. \tag{A.2.2}$$

The parameter $\delta$ is a time-domain analogue of $R$, representing the time window over which the sinusoid velocity-reversal component acts, as per the piece-wise component limits in Eq. A.2.1; and $T$ is the waveform period, $T = 2\pi/\omega$. Analytical formulations of $x(t)$, via Eq. A.2.2, are available, but these formulations are not required for frequency-band acceleration-matching, and so we use numerical integration for convenience.